\documentclass[A4 paper, 10pt]{article}

\usepackage[T1]{fontenc}
\usepackage{latexsym}
\usepackage{physics}
\usepackage{amsmath}
\usepackage{scrextend}
\usepackage{authblk}
\usepackage{amssymb}
\usepackage{slashed}
\usepackage{empheq}
\usepackage[english]{babel}
\usepackage{booktabs} 
\usepackage{array}
\usepackage{verbatim} 
\usepackage{xcolor}
\usepackage{soul}
\usepackage[pdfa]{hyperref}
\usepackage[]{pdfpages}
\usepackage{graphicx}

\usepackage[title]{appendix}
\usepackage{subcaption}
\usepackage{mathtools}
\usepackage{float}
\newtagform{Eq}[\upshape]{[}{]}
\usetagform{Eq}
\let\vec\mathbf

\begin{document}
	\title{Back to the Roots of Vector and Tensor Calculus.\\ Heaviside versus Gibbs.}

	\author{Alessio Rocci\thanks{a\_rocci@hotmail.com}}
	\affil[1]{Department of Physics and Astronomy ``G. Galilei'', via Marzolo 8, I-35131 Padova (Italy)}
	\maketitle

\begin{abstract}
In June 1888, Oliver Heaviside received by mail an officially unpublished pamphlet, which was written and printed by the American author Willard J. Gibbs around 1881-1884. This original document is preserved in the Dibner Library of the History of Science and Technology at the Smithsonian Institute in Washington DC. Heaviside studied Gibbs's work very carefully and wrote some annotations in the margins of the booklet. He was a strong defender of Gibbs's work on vector analysis against quaternionists, even if he criticized Gibbs's notation system. The aim of our paper is to analyse Heaviside's annotations and to investigate the role played by the American physicist in the development of Heaviside's work.
\end{abstract}

\tableofcontents

\section{Introduction}

The roots of modern vector and tensor calculus go back to the end of the nineteenth century, when two authors independently developed the modern system \cite{Crowe}. The two lived and worked on opposite sides of the Atlantic Ocean, Willard J. Gibbs was an American physicist, while Oliver Heaviside was a British scientist. Until the end of the 1880s, they worked separately using similar mathematical tools, but in 1888 Gibbs sent a copy of his unprinted pamphlet \textit{Elements of vector analysis} to Heaviside and to many other scientists, in order to promote the use of vectors in physics. Heaviside received Gibbs's booklet in June and Gibbs received Heaviside's answer in July, in which the British scientist acknowledged receiving reprints of Gibbs's lectures (\cite{Gibbs-biblio}, 224). Heaviside realized that he was not the only one working in that area. He was impressed by Gibbs's work and recognized that the American physicist had reached some useful advances in the area. Both authors predicted that vectors would play an important role in theoretical physics, as indeed they did. 

Both authors began to develop vector calculus after studying Maxwell's \textit{Treatise on electricity and magnetism} and both were impressed by Maxwell's criticism of the quaternion system, which he tried to apply for describing electromagnetic phenomena. Both Gibbs and Heaviside knew quaternions, an abstract system discovered by Sir Rowan Hamilton in the 1830s \cite{Hamilton} \cite{Crowe} where scalar and vector quantities can be added together, analogous to the sum between real and imaginary quantities in the context of complex numbers. Like Ari Ben-Menahem observed, `Maxwell's work made clear that vectors are a real tool for physical thinking and not just an abbreviated notation' (\cite{Ben-Menahem-Encyclopedia-3}, 1814) and in Maxwell's theory, scalars and vectors had different and separate roles. In Heaviside's words: `once a vector always a vector' (\cite{EM1}, 301). And the same is true for scalar quantities. Hence, both scientists developed a system where scalars are summed with scalars only, and vectors with vectors only. Unlike Heaviside, the American physicist was attacked by Peter G. Tait in the preface of his book on quaternions and was accused of lengthening the diffusion of the quaternionic system in physics \cite{Tait-3rd}. In this public debate, Heaviside defended Gibbs's position (\cite{EM2}, 557) (\cite{EM1}, 137), though the American physicist did not need any help in answering in \textit{Nature} journal \cite{Nature-Gibbs}, where the dispute continued \cite{Nature-Tait}, also in the following years\footnote{See \cite{Crowe} for a review of the debate.}.

As said above, Heaviside had great respect for Gibbs's work, although he did not like, and sometimes criticized, Gibbs's choices for vector notation. 
Bruce Hunt reported that `Heaviside was a self-trained English mathematical
physicist [...] [he] spent his career on the far fringes of the scientific
community' (\cite{Hunt}, 48). Paul Nahin confirmed that Heaviside was used to working by himself or with his brother for most of his career \cite{Heav-biblio-Nahin}. The purpose of our paper is to shed light on the connections between Gibbs and Heaviside. In order to investigate the impact that Gibbs's approach had on Heaviside's work, we analysed the booklet sent by Gibbs to Heaviside in 1888, which is now preserved at The Dibner Library of the History of Science and Technology. Indeed, the pamphlet contains many manuscript annotations, which suggest that Heaviside studied and analysed Gibbs's booklet thoroughly. Our paper is organized as follows. 

In section \ref{Gen-remarks}, we present some general remarks. First, we argue on the authenticity of the annotations by comparing them with different Heaviside manuscripts held at the Dibner. Second, we address the problem of dating the annotations. We shall propose a time interval we inferred by comparing Heavside's annotations with his published work. Third, we review the structure of Gibbs's pamphlet and we  present some concepts of vector and tensor calculus that are important for the purpose of our paper. In this context, we shall introduce and discuss a unifying notation we chose to compare Gibbs's and Heaviside's results.

In the rest of the paper, we shall analyse the annotations comparing them with Heaviside's published works, the \textit{Electrical Papers} \cite{ELP1} \cite{ELP2}, held in the Dibner Library, and the \textit{Electromagnetic Theory} \cite{EM1}, \cite{EM2} and \cite{EM3}. In the main text, we shall present what we considered the most important annotations, because they represent the origin of some reflection we found in Heaviside's published work. For completeness, the remaining, more ordinary, annotations, which we consider less important, will be analysed in an appendix.

The annotations presented in our main text are organized as follows. First, we shall analyse the annotations that represent the origin of Heaviside's discussions on concepts he had already considered. In section \ref{Heavi-vector-annot}, we shall explain how Heaviside discussed the definition of the reciprocal of a vector and how he reflected on the abstract concept of ``product'' in mathematics by discussing Gibbs's indeterminate product. Second, in the following sections, we shall analyse the annotations that originated Heaviside's reflections on new concepts, especially in the context of linear operators. In section \ref{lin-operator-section}, we shall present Heaviside's triadic, i.e. a particular tensor of rank three in modern language, which does not appear in Gibbs's booklet. The section ends with an analysis of Heaviside's reflections on the physical meaning of the curl of a tensor. 

During the last years of the nineteenth century, after having received Gibbs's booklet, Heaviside elaborated a generalisation both of the Divergence theorem and of the Stokes's theorem, an issue already partially discussed by Ido Yavetz \cite{Heav-biblio-Yavez}. In section \ref{Green-Stokes}, we shall present a formula for transforming surface-integrals into line-integrals that Heaviside published in his \textit{Electromagnetic Theory} in 1893. Both Gibbs and Heaviside presented their own generalisation of the two theorems. The two formulations appear to be very different, therefore, we tried to address the following questions. Is there any connection between Gibbs and Heaviside's generalisations of the Divergence and the Stokes's theorem? Did Heaviside use some of the concepts described in Gibbs's pamphlet? In section \ref{Green-Stokes}, we argue that Gibbs's booklet stimulated Heaviside's creativity, by comparing the annotations with the published generalisation of the two theorems.

The paper ends with section \ref{Conclusions} and three appendices,  \ref{parity}, \ref{app1} and \ref{others}. In section \ref{Conclusions}, we shall summarize our analysis of Heaviside's annotations, and then, in appendix \ref{parity}, we shall discuss technical details regarding the connection between Gibbs and Heaviside's generalisations of the Divergence theorem, which we skipped in section \ref{Green-Stokes}. In appendix \ref{app1}, we shall present a complete transcription of the annotations on the back cover of the booklet, which is the most damaged part of the pamphlet, and in appendix \ref{others}, we present the ordinary annotations we did not discuss in the main text.

\section{General remarks}\label{Gen-remarks}
\subsection{Proof of authenticity}
In 1888 Heaviside received by mail a copy of a pamphlet on vector analysis written by Gibbs \cite{Gibbs-pamphlet}. Gibbs had sent it to him and to many other scientists, journals and institutions (\cite{Gibbs-biblio}, 247 and \cite{Crowe},154). Gibbs's booklet remained unpublished until Edwin B. Wilson, one of Gibbs's student, was asked to write a book based on Gibbs's pamphlet, broadening its subject \cite{Wilson}. Heaviside himself expressed disappointment about the fact that Gibbs's booklet remained unpublished (\cite{ELP1}, 529). We know directly from Heaviside that he received a copy of the pamphlet in June 1888 (\cite{ELP2}, 529). In this section, we shall argue that the booklet held in the Dibner Library is the copy received by Heaviside. 

An annotation written in pen appears on the front cover of the booklet: `From the author. June 1888.' (\cite{Gibbs-pamphlet}; front cover page). This suggests the first evidence of the authenticity of the pamphlet. On the front cover, there is also a brief annotation, written in pencil: ``Mss notes by Oliver Heaviside and on back of cover'', where \textit{Mss notes} means manuscript notes. 

\begin{figure}[ht!]
	\begin{subfigure}{.5\textwidth}
		\centering
		\includegraphics[width=40mm]{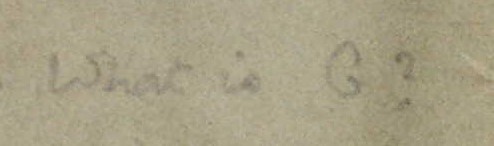}
		\caption{Gibbs's booklet}
		\end{subfigure}%
	\begin{subfigure}{.5\textwidth}
		\centering
		\includegraphics[width=35mm]{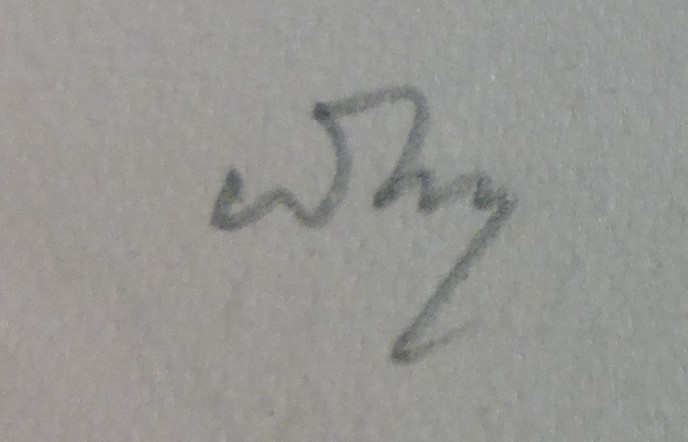}
		\caption{Heaviside's manuscript}
		\end{subfigure}
	\caption{\textbf{Left:} `What is G?'. \textbf{Right:} `Why'. Cfr. letter W and h in the two pictures: the two are attached to each other in both pictures}
	\label{W}
\end{figure}

\begin{figure}[ht!]
	\begin{subfigure}{.5\textwidth}
		\centering
		\includegraphics[width=28mm]{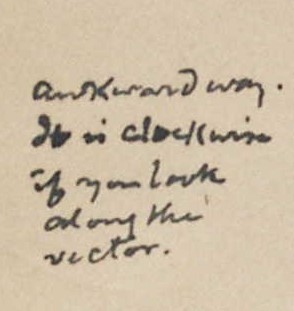}
		\caption{Gibbs's booklet}
	\end{subfigure}%
	\begin{subfigure}{.5\textwidth}
		\centering
		\includegraphics[width=35mm]{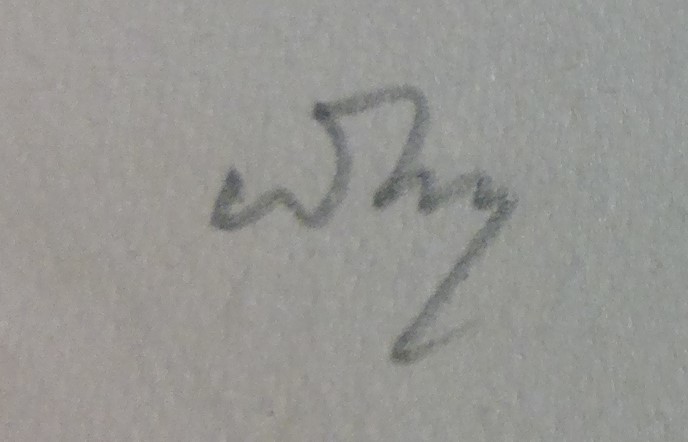}
		\caption{Heaviside's manuscript}
	\end{subfigure}
	\caption{\textbf{Left:} third line, `if you look'. \textbf{Right:} `Why'. Cfr. letter y in the two pictures}
	\label{y}
\end{figure}

\begin{figure}[H]
	\begin{subfigure}{.5\textwidth}
		\centering
		\includegraphics[width=35mm]{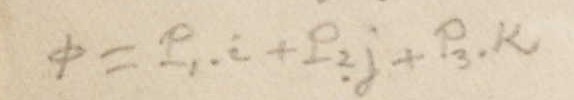}
		\caption{Gibbs's booklet}
	\end{subfigure}%
	\begin{subfigure}{.5\textwidth}
		\centering
		\includegraphics[width=25mm]{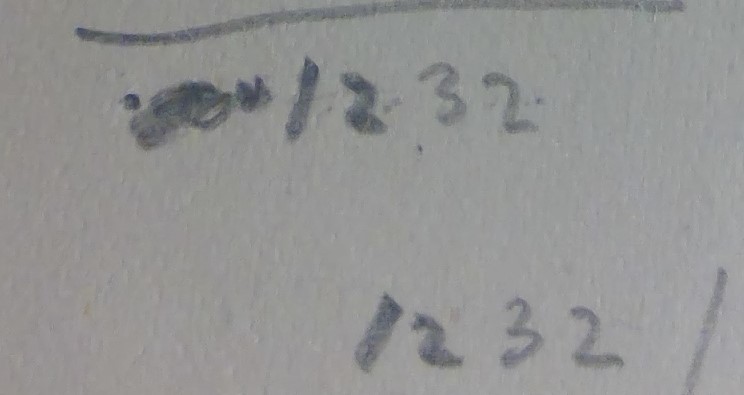}
		\caption{Heaviside's manuscript}
	\end{subfigure}
	\caption{Cfr. the shape of the numbers in both pictures}
	\label{numbers}
\end{figure}

\begin{figure}[ht!]
	\begin{subfigure}{.5\textwidth}
		\centering
		\includegraphics[width=28mm]{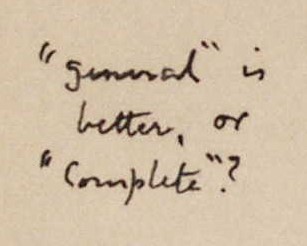}
		\caption{Gibbs's booklet}
	\end{subfigure}%
	\begin{subfigure}{.5\textwidth}
		\centering
		\includegraphics[width=35mm]{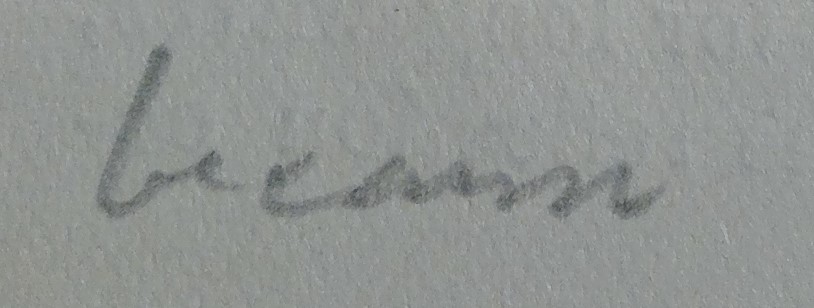}
		\caption{Heaviside's manuscript}
	\end{subfigure}
	\caption{\textbf{Left:} third line, `better, or'. \textbf{Right:} `because', Cfr. letter b in the two pictures}
	\label{b}
\end{figure}

The pamphlet contains marginal annotations both in pencil and in pen, which seem to have been written by the same hand, as reported in the catalogue description of the Dibner Library's website. First, we verified the authenticity of the handwriting. For this purpose, we used manuscripts signed by Heaviside himself (MSS 677 A) also held at the Dibner Library \cite{Heaviside-MS1}. Using these notes, we compared the calligraphy seeking matching letters or numbers. Indeed, we found at least three pieces of evidence: the shape of the letters \textit{W} and \textit{h} (Fig. \ref{W}), that of letter \textit{y} (Fig. \ref{y}), the numbers \textit{1}, \textit{2} and \textit{3} (Fig. \ref{numbers}) and the shape of letter \textit{b} (Fig. \ref{b}). In the following pictures, the two different sources are compared: on the right side MSS 677 A, while on the left side some of the annotations on Gibbs's pamphlet. All the photographs of our paper are provided by kind permission of the Dibner Library.

Having established the authenticity of the annotations, we investigated the difference between the two types of annotations. The ones written in pencil are less in number, are always placed close to the text they refer to and  mark the presence of a formula. They have no special role. The annotations written in pen have a different role. There are brief comments and many formulas. The former are reflections on Gibbs's text, the latter are both ``translations''\footnote{Our use of the term ``translation'' will be further clarified in section \ref{translation}} of Gibbs's formulas in Heaviside's language and developments based on Gibbs's formulas. By analysing the annotations written in pen, we found some specific terms that Gibbs would introduce later in the text, i.e. the terms \textit{dyad} and \textit{dyadics}. As far as we know, Heaviside had never used the term \textit{dyadic} in his work before receiving Gibbs's pamphlet. Hence, we inferred that the annotations written in pen were made after Heaviside had read the whole booklet.

\subsection{Dating the annotations}\label{dating}
Another important question is: when did Heaviside write his annotations? We do not have access to any of Heaviside's personal diaries, if they exist, therefore the holographs cannot be dated exactly. The best we can do is give the shortest lapse of time. Even if it is obvious that the annotations were made after June 1888, it is far less obvious before which date they were made. In a footnote on Heaviside's Electrical Papers, commenting Gibbs's work he had received, the author asserted: `It is indeed odd that the author should not have published what he had been at the trouble of having printed. \textit{His treatment of the linear vector-operator is especially deserving of notice.}' [emphasis added] (\cite{ELP2}, 529). Hence, we can infer that, at the time of preparing the proofs of the Electrical Papers, Heaviside had read the whole of Gibbs's pamphlet, because in the footnote he referred to the linear operators, i.e. Gibbs's dyadics. More precisely, the footnote has been written after November 13, 1891, as can be inferred with further reading of the footnote itself. In the same period, a series of papers by Heaviside on vector analysis started to be published on \textit{The Electrician}. Those papers were subsequently published as the third chapter in the first volume of Heaviside's work \textit{Electromagnetic Theory}. Here the term dyadic appeared for the first time in Heaviside's work when the author introduced linear operators: `This is (with changed notation, however) Prof. Gibbs's way of regarding linear operators. The arrangement of vectors in (26) [which] he terms a \textit{dyadic}, each term of two paired vectors [...] being a \textit{dyad}.' [emphasis added] (\cite{EM1}, 263). This quotation was published on September 2, 1892, in \textit{The Electrician}. 

Further evidence of the fact that Heaviside had studied the booklet in 1892 is the following. After the last quoted sentence, he remarked that `Prof. Gibbs has considerably developed the theory of dyadics.' (\cite{EM1}, 263). Furthermore, on April 8, 1892, in his first summary on vector algebra, a particular formula\footnote{The formula will be discussed in section \ref{Green-Stokes}.} was published by \textit{The Electrician}, and the same formula appears in one of Heaviside's annotations in Gibbs's pamphlet. Furthermore, as we shall discuss in section \ref{reciprocal}, in June 1892 Heaviside used Gibbs's notation for linear operators, which he would criticize in his following work.

Finally, on page 43, Heaviside translated Gibbs's expressions for the vector product between a vector and a dyadic, i.e. a linear operator. 
\begin{figure}[ht!]
	\centering
	\includegraphics[width=90mm]{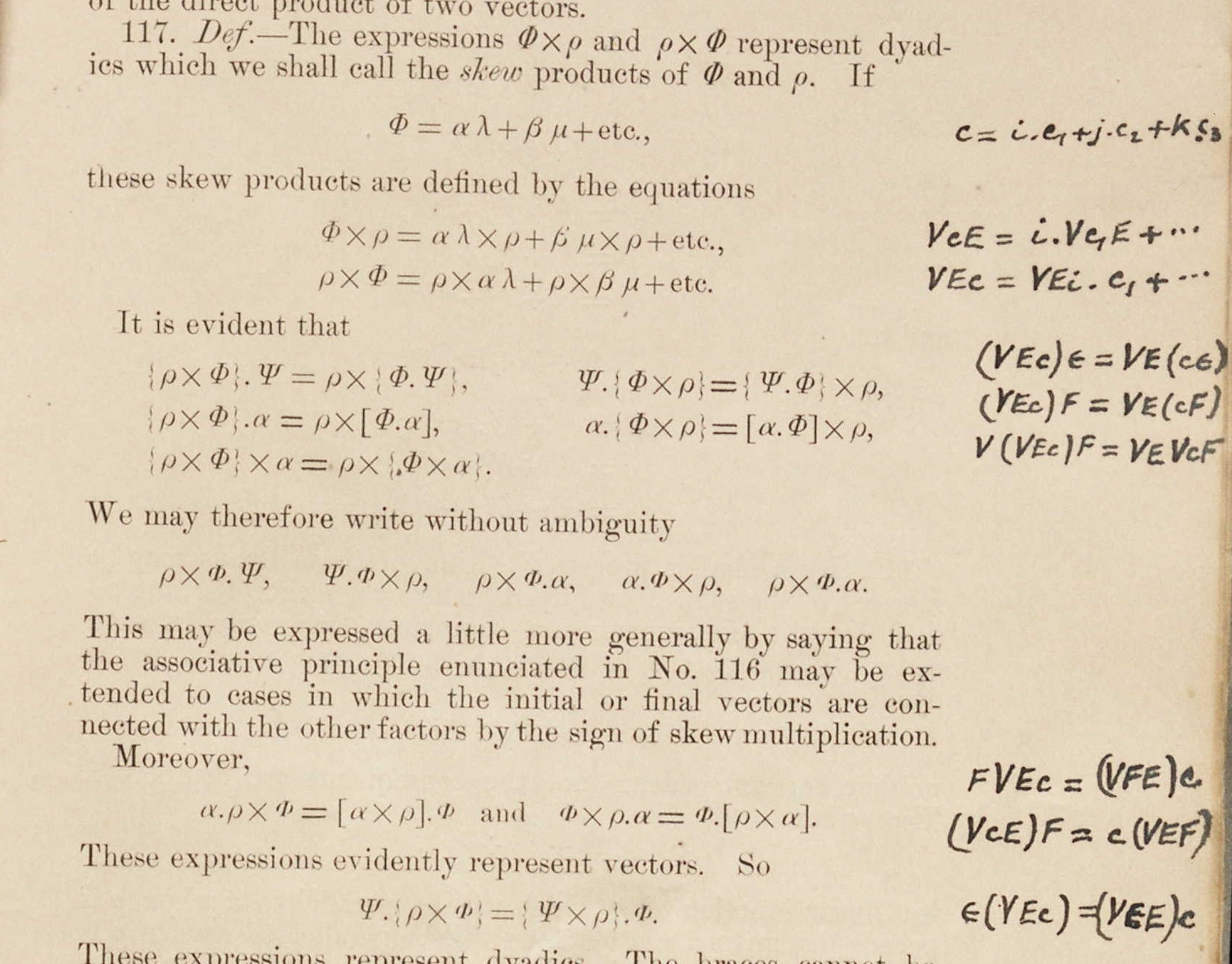}
	\caption{\textbf{First three lines:} these equations appear also in \textit{Electromagnetic Theory}. The letter \textit{c} represents a linear operator, while $ \vec{E} $ is a vector. \textbf{Other lines:} various products involving the linear operator $ c $ and two vectors $ \vec{E} $ and $ \vec{F} $ (p. 43)}
	\label{vector-product-dyadic}
\end{figure}
The letters used by Heaviside are very similar both to those he used in the 1892 note he added to the published edition of the \textit{Electrical Papers} (\cite{ELP2}, 19) and to those that appear in \textit{Electromagnetic Theory} (\cite{EM1}, 262). In both works, Heaviside emphasised the importance of Gibbs's contribution to the theory of linear operators. In \textit{Electromagnetic Theory}, the formulas of Fig. \ref{vector-product-dyadic} finally appeared (\cite{EM1}, 283-286-295).

 Hence, we can infer that Heaviside studied Gibbs's booklet between June 1888 and April 1892 and that the annotations were made during this period.

\subsection{The structure of Gibbs's work and a unifying notation}\label{VecReview}
Before proceeding, the terms \textit{dyad} and \textit{dyadic} should be clarified, because they will appear again. As said above, Gibbs and Heaviside used different notations. In section \ref{Heavi-vector-annot}, we shall present both notations, but in order to compare them, we decided to use a unifying modern notation, which we introduce in the following. 

Let us start by analysing the structure of Gibbs's work. In chapter one, entitled ``Concerning the algebra of vectors'', the author introduced some fundamental notions, the sum and the products of vectors with their properties. The chapter concludes with a treatment of methods for solving vectorial equations. In chapter two, entitled ``Concerning the differential and integral calculus of vectors'', Gibbs introduced the \textit{nabla} operator, denoted by the symbol $ \nabla $, and its applications, i.e. the mathematics of potential theory. After having introduced line, surface and volume integrals, Gibbs presented the Stokes's theorem and the Divergence theorem for vector fields. The chapter ends with the introduction of irrotational and solenoidal vector fields and with a brief analysis of the infinities arising in the context of volume integrals. These two chapters were printed in 1881, while the remainder of the text in 1884. In chapter three, entitled ``Concerning linear vector functions'', Gibbs introduces the reader to the concept of linear operators, by defining the terms \textit{dyad} and \textit{dyadics}. As an application of this concept, Gibbs analysed rotations and strains. Chapter four is entitled like chapter two and it presents some supplementary material. In this chapter, Gibbs considers the divergence and the curl of a linear operator and presents a generalisation of the Stokes's theorem and the Divergence theorem for linear operators. Chapter five is dedicated to transcendental functions of dyadics, while the last chapter, entitled ``Note on bivector analysis'', is dedicated to the algebra of complex-valued vectors.   

Gibbs defined vectors as in modern textbooks: `If anything has magnitude and direction, its magnitude and direction taken together constitute what is called a vector.' (\cite{Gibbs-pamphlet}, 1). Given two vectors, he defined three kinds of products, each of them resulting in three different mathematical objects. From a modern point of view, the three types of products described in Gibbs's booklet are the scalar product, the vector product and the tensor product. The latter was called by Gibbs the ``indeterminate product'' and we shall discuss his choice of the name in section \ref{indeterminate}. Heaviside used the term ``tensor'' but with a different meaning. In Heaviside's work, ``the tensor of a vector'' is the magnitude of the vector itself. As already said, Heaviside derived his notation from the school of quaternionists: Hamilton introduced the term \textit{tensor of a quaternion} in \cite{Hamilton}, by defining it as the generalisation of the modulus of a complex number. Hamilton himself referred to the intensity of a vector, which is a particular quaternion, as ``the tensor of the vector''.

Even nowadays, there is no universal notation for identifying vector objects. Frequently used notations for vector quantities are an arrow or a hat as a superscript over the letter representing them or writing Latin letters in bold script. As quoted by Heaviside himself, this last notation was introduced by Clarendon in August 1886 in a paper published in the \textit{Philosophical Magazine} and it was adopted also by the English author (\cite{ELP1}, 199). Gibbs used Greek letters instead. In our paper, it is not possible to use a single notation for denoting vectors. We will use Clarendon's in this section. In section \ref{Heavi-vector-annot}, describing Heaviside's annotations on vector algebra, we will maintain the original notation introduced by Gibbs, i.e. Greek letters. Finally, we will compare Heaviside's annotations with his published papers using both Clarendon's notation and Greek letters. 

Also the symbols used by Gibbs and Heaviside for the products were different and, again, also the notation used nowadays is far from unique. Vectors notation simplified considerably many formulas in physics because it allows to avoid the use of Cartesian coordinates. From this perspective, the differential absolute calculus played a similar role for linear operators. Indeed, as it is well known, the term absolute means that its equations are formal invariants for arbitrary changes of coordinates. Therefore, in order to compare and clarify the relationship between Gibbs and Heaviside's notation, we decided to introduce the index notation as unifying notation. Like in Heaviside's annotations, we shall translate both notations both for clarity and simplicity and for historical reasons. Indeed, we are not the first to translate Gibbs's notation into the language of absolute differential calculus. In 1926, soon after the introduction of the generalised Kronecker symbol and of its connection with the Levi-Civita alternating symbol \cite{Murnaghan}, Clarence L. E. Moore translated, as far as we know for the first time, Gibbs's indeterminate product (and dyadics) as well as Grassmanian inner and outer product in order to show how all products can be described using the concept of index-contraction \cite{Moore}. 

Given an arbitrary vector $ \vec{a}$ (Clarendon notation), let $ a_i $ be its real components with respect to an arbitrary basis, where Latin indices range in the set $ {1\, ,\,2\, ,\,  3} $. The three products, using vector notation and index notation, read:
\begin{eqnarray}
\text{scalar product:} &\qquad & \vec{a} \boldsymbol{\boldsymbol{\cdot}}\vec{b} \;\,\qquad a_ib_i\\
\text{vector product:} &\qquad & \vec{a}\times\vec{b}\qquad \epsilon_{ijk}a_ib_j\\
\text{tensor product:} &\qquad & \vec{a}\otimes\vec{b}\qquad a_ib_j\; .\label{tens-def-1}
\end{eqnarray}
We adopted Einstein's summation rule \cite{summation} and the Levi-Civita symbol $ \epsilon_{ijk} $ is defined by $ \epsilon_{123} = 1 $.  

In Gibbs's booklet, the tensor product of two vectors is called a \textit{dyad} and it represents a linear operator $ \Psi =  \vec{a}\otimes\vec{b} $. In general, linear operators are maps between two different vector spaces, but neither Gibbs nor Heaviside specified this fact, because in the applications they investigated, the two authors considered linear mapping from a vector space to itself only. A dyad can act on a vector $ \vec{r} $ either on the right or on the left. It acts on the right as follows: $ \Psi\boldsymbol{\cdot}\vec{r} \overset{def}{=} \vec{a}\left( \vec{b}\boldsymbol{\cdot}\vec{r}\right)  $, while the action on the left can be defined analogously. Using index notation, $ \Psi $ is represented by a $ 3\times 3 $ matrix, namely $ \Psi_{ij}\overset{def}{=} a_ib_j $. Therefore, dyads can be identified with special tensors because a generic element of the vector space $ V\otimes V $ cannot always be written as the tensor product of two vectors\footnote{As usual, we used the same symbol for denoting both the tensor product of vector spaces and of vectors, but the former was introduced later than the latter. The tensor product of vector spaces had been formally introduced in the nineteenth century and, as far as we know, the symbol $ \otimes $ appeared for the first time in this context \cite{Neumann}.}. Using modern language, they correspond to \textit{simple}, or elementary, tensors (\cite{I-Shih}).

In Gibbs's language, a linear combination of dyads is a dyadic. A dyadic constructed using the six vectors $ \vec{a} $, $ \vec{b} $, $ \vec{c} $, $ \vec{l} $, $ \vec{m} $ and $ \vec{n} $ is: $  \Phi = \vec{a}\otimes\vec{l} + \vec{b}\otimes\vec{m} + \vec{c}\otimes\vec{n} $ and it represents the more general form of a linear operator. Using index notation $ \Phi $ reads $ \Phi_{ij}\overset{def}{=} a_il_j + b_im_j +c_in_j $. In the application considered by Heaviside and Gibbs, dyadics can be identified with generic tensors of rank two\footnote{In the following, also triadics will appear, which will correspond to tensors of rank three.}.

In the chapter dedicated to stress and strain, Gibbs considered the linear operators in the context of continuum mechanics. A dyadic $ \Phi $ would describe an action on the body like a deformation or a rotation. As emphasised in \cite{Wilson}, Gibbs denoted a \textit{generic} set of orthonormal vectors, not necessarily the canonical basis like nowadays, with $ \vec{i} $, $ \vec{j} $, $ \vec{k} $. In the following, two special dyadics will appear. The first dyadic, namely
\begin{equation}\label{id-dyadic}
	\mathcal{I}=\vec{i}\otimes\vec{i} + \vec{j}\otimes\vec{j}+ \vec{k}\otimes\vec{k}\; ,
\end{equation}
is represented in components by the Kronecker delta $ \delta_{ij} $ (\cite{Moore}, 199), and maps every vector into itself. When a generic orthonormal set is mapped into a different orthonormal set, namely $ \vec{i}' $, $ \vec{j}' $, $ \vec{k}' $, Gibbs introduced the following dyadic:
\begin{equation}\label{rotation-matrix}
\mathcal{R}=\vec{i}'\otimes\vec{i} + \vec{j}'\otimes\vec{j}+ \vec{k}'\otimes\vec{k}\; ,
\end{equation}
but he did not use a single letter for denoting dyadic (\ref{rotation-matrix}) in his booklet, we chose $ \mathcal{R} $ because, as specified by Gibbs himself, it represents a rotation.

\section{Heaviside's annotations on vector algebra}\label{Heavi-vector-annot}
\subsection{Comparing notations}\label{translation}
As Gibbs had prophesied, a vigorous debate on vectorial methods took place at the end of the 1890s (\cite{Gibbs-biblio}; 115). One of the main themes debated was the particular notation adopted for the operations between vectors and scalars. The modern system of vector algebra emerged after the discovery of quaternions, as widely investigated by \cite{Crowe}. Looking for a three-dimensional generalisation of complex numbers, Sir Rowan Hamilton introduced the quaternion system\footnote{See \cite{Tait-1st} for an introduction.} in 1843. In 1873, Maxwell wrote his equations both in components and in quaternion notation (\cite{Crowe}, 128). He studied Hamilton's system, but he was critical of the utility of quaternions. Gibbs and Heaviside developed independently the vector calculus in order to find a more suitable tool for representing three-dimensional physical quantities, for the electromagnetic theory and for different areas. The different notations used by the two authors are also explained by Crowe, who analysed how Heaviside adhered, with slight differences, to the quaternionist notation, well represented by Peter G. Tait and by his Treatise \cite{Tait-1st}, while Gibbs invented a new notation. The differences between the two authors in vector algebra are the following:
\begin{eqnarray}
\quad\quad&\text{Heaviside}&\quad\quad \text{Gibbs} \nonumber \\
\text{scalar product}\quad\quad &\vec{a}\vec{b} &\quad\quad\;\, \alpha .\beta \\
\text{vector product}\quad\quad &V\vec{a}\vec{b} &\quad\quad \alpha\times\beta\\
\text{tensor product}\quad\quad & \vec{a}.\vec{b} & \quad\quad\;\;\alpha\beta\qquad .
\end{eqnarray}
Heaviside adopted Clarendon's notation with Latin letters, while Gibbs used Greek letters only. Heaviside used the period as a mere separator, while Gibbs introduced it as a symbol for the scalar product. The period became a dot like the one used today in Wilson's book (\cite{Wilson}, 55). In order to avoid confusion between Heaviside and Gibb's period, we shall write $ \alpha\boldsymbol{\cdot}\beta $ instead of $ \alpha .\beta \,$ for the scalar product in the main text as well as in Gibbs's quotations.

The use of different notations reflected the different philosophical approach the author had toward Mathematics. The three products obey to different properties. For example, only the scalar product is commutative. The symbols used by the two authors reflected their different attitude with respect to this fact. Heaviside was more conservative than Gibbs. Indeed, as he himself remarked, he decided that it would have been too ugly to introduce a symbol between vectors for the scalar product, because it obeys the same laws as the product between numbers, e.g. commutativity law. Heaviside himself remarked the differences between his and Gibbs's notation in a footnote he added to a previous work republished in \textit{Electromagnetic Theory}: `In short, I amalgamate the members of products; Gibbs separates them.' (\cite{EM3}, 142). Heaviside adopted the quaternion notation for the vector product by using the capital letter $ V $ before the cross product between two vectors. We don't know why Gibbs invented his own notation, but his attitude was more modern because nowadays we introduce new symbols for operations between new objects. Indeed, Gibbs's notation and Gibbs's philosophy survived, for reasons not fully explored yet. If not otherwise stated, we will use modern symbols for scalar and vector products and, for denoting vector quantities, both Greeks letters and bold Latin letters, e.g. $ \vec{a}\cdot\vec{b} $ and $ \vec{a}\times\vec{b} $ or $ \alpha\cdot\beta $ and $ \alpha\times\beta $.

Heaviside started to use vectors independently in 1882 (\cite{ELP1}, 199). When he received Gibbs's booklet he had used identities for the double vector product $ (\vec{a}\times\vec{b})\times\vec{c}  $ and for the mixed product, i.e. $ (\vec{a}\times\vec{b})\boldsymbol{\cdot}\vec{c}  $, and he had also used the \textit{nabla} operator in order to introduce the gradient of a scalar function $ f $, namely $ \nabla f $, the \textit{curl} and the divergence of a vector $ \vec{a} $, denoted with $ curl\,\vec{a} $ and $ div\,\vec{a} $ respectively. Heaviside had introduced vectors for describing electromagnetic phenomena and in the context of continuum mechanics. As already said, Heaviside did not like Gibbs's notation for vector algebra. Hence, it is not surprising that, in most of his annotations, Heaviside rewrote Gibbs's equations using his own notation. In the following, we shall use the verb ``to translate'' for denoting this fact. It is worth noting that in 1888 Heaviside had not used, in his published papers, all of the vector identities presented in Gibbs's booklet and that he had not written his first summary on vector algebra yet, whose publication started in November 1891  (\cite{EM1}, 132). Therefore, by analysing Heaviside's annotations, we shall address the following questions: What was Heaviside's approach in analysing Gibbs's pamphlet? Is there any evidence of Gibbs's influence on Heaviside's work?

\begin{figure}[H]
	\centering
	\includegraphics[width=90mm]{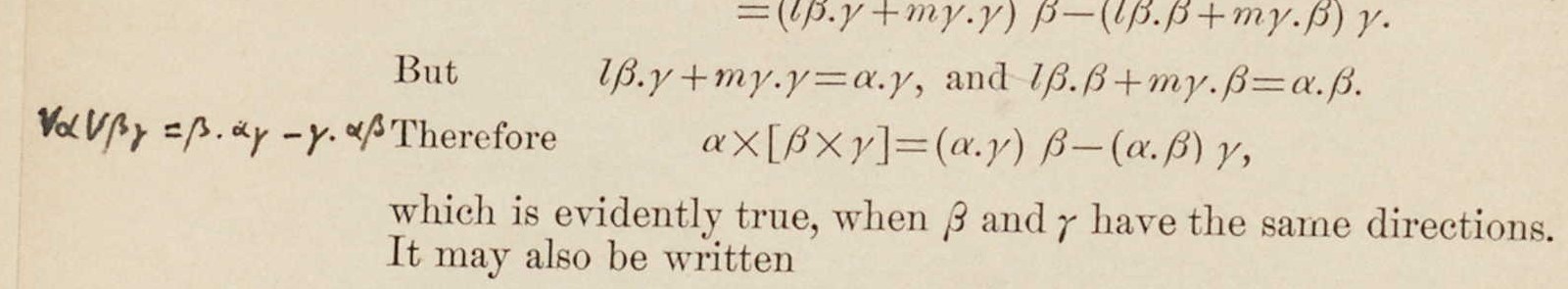}
	\caption{Double vector product formula translated (p. 8)}
	\label{double-vector-product}
\end{figure}

First, we present a typical example of Heaviside's translation, i.e. his first annotations on page 8. Heaviside translated the double vector product formula without discussing it\footnote{There is also a note written in pencil, which is identical to that written in pen.}, Fig. \ref{double-vector-product}. In his translation, he used his symbols for the vector and the scalar product, but he maintained Gibbs's notation, i.e. the Greek letters, for denoting vectors:
\begin{eqnarray}
&\text{Gibbs}&\quad\quad\qquad\qquad \text{Heaviside} \nonumber \\
&\alpha\times\left[ \beta\times\gamma\right]  = (\alpha \boldsymbol{\cdot} \gamma)\beta - (\alpha \boldsymbol{\cdot}\beta)\gamma &\quad\quad\quad\, V\alpha V \beta\gamma = \beta . \alpha\gamma - \gamma .\alpha\beta
\end{eqnarray}
It is worth noting that in Heaviside's notation the parentheses are not needed, as he himself emphasised in \textit{Electromagnetic Theory}.

\begin{figure}[ht!]
	\centering
	\includegraphics[width=70mm]{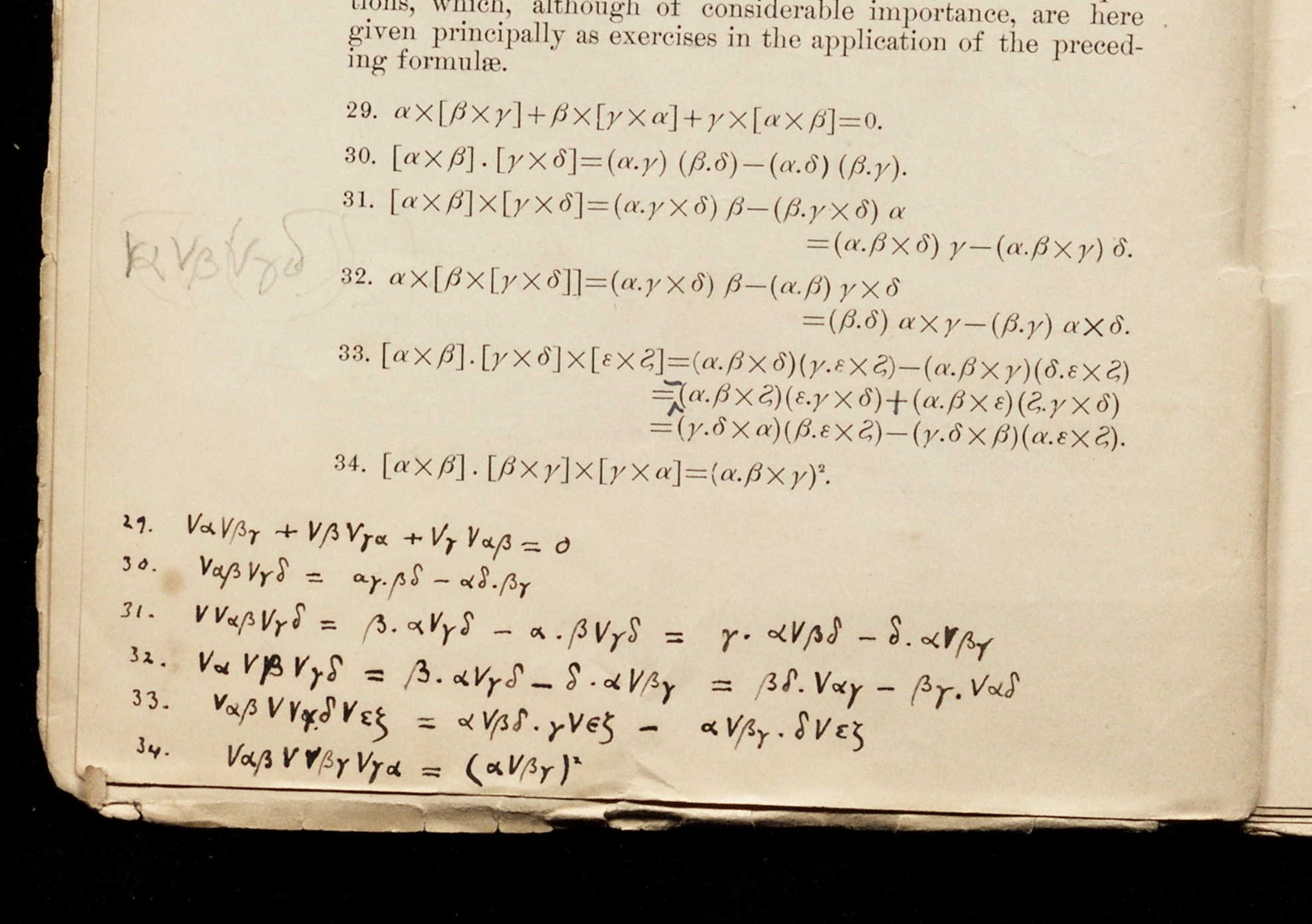}
	\caption{Heaviside's translations on page 8. Eq. (29) corresponds to the cyclic property of the cross product}
	\label{translations-p-8}
\end{figure}

The second example is a list of translations at the bottom of page 8, Fig. \ref{translations-p-8}. The formulas are identified with the same number used by Gibbs. In order to avoid confusion, the equations of our paper will be denoted by squared brackets, e.g. eq. [9], while equations referring to Gibbs's booklet or Heaviside's papers are denoted by using the usual parentheses, e.g. eq. (29) in Fig. \ref{translations-p-8}. 

\subsection{On the \textit{reciprocal} of a vector}\label{reciprocal}
As we emphasised in the introduction, some of Heaviside's critiques on Gibbs's approach can be traced back to the annotations of the booklet. First, we shall present two examples regarding the possibility of defining the reciprocal of a vector, a concept that Heaviside had already introduced.  

\begin{figure}[ht!]
	\centering
	\includegraphics[width=70mm]{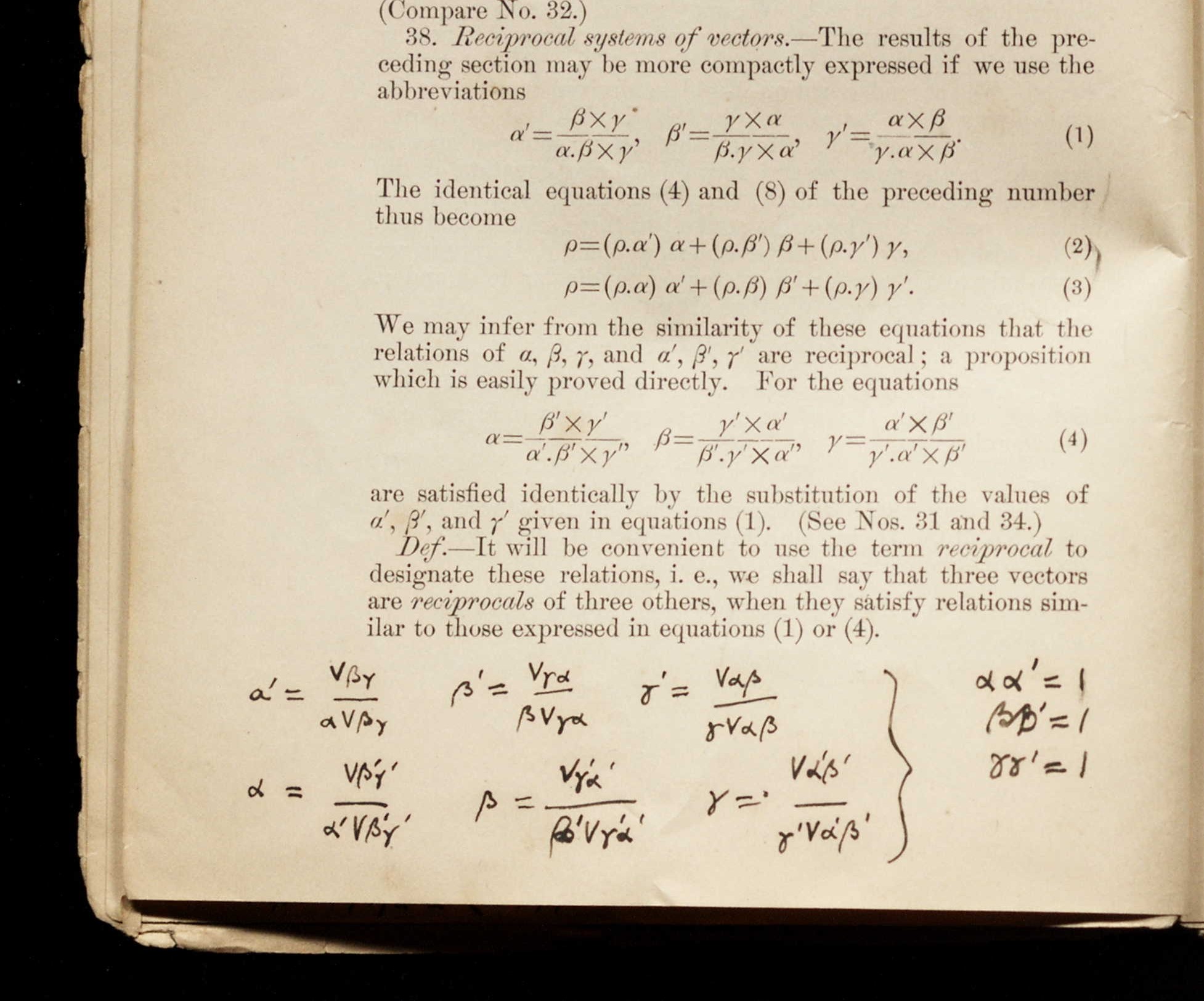}
	\caption{Heaviside's translation of the reciprocal system of vectors (p. 10)}
	\label{reciprocal1}
\end{figure}

On page 10, see Fig. \ref{reciprocal1}, Heaviside translated both the formulas for obtaining what Gibbs called `the reciprocal system'\footnote{In modern language, by using the component notation, when the vectors of a basis are represented by upper indices $ \vec{e}^i $, the reciprocal system corresponds to the \textit{dual} basis, denoted by lower indices, because $ \vec{e}^i\vec{e}_j = \delta^i_j $, where the Kronecker symbol has the same properties as $ \delta_{ij} $. The term reciprocal survived until today in the context of crystallography.} $ \alpha'$,  $ \beta'$,  $\gamma' $ of a given set of three non-coplanar vectors $ \alpha $, $ \beta$,  $ \gamma $ (\cite{Gibbs-pamphlet}, 10) and the conditions defining the reciprocal system, namely $ \alpha\boldsymbol{\cdot}\alpha' =\beta\boldsymbol{\cdot}\beta'=\gamma\boldsymbol{\cdot}\gamma' = 1 $ and $ \alpha\boldsymbol{\cdot}\beta' =\gamma\boldsymbol{\cdot}\beta'=\alpha\boldsymbol{\cdot}\gamma' = 0 $. 
The condition $ \alpha\cdot\alpha' = 1 $ could suggest defining $ \alpha' $ as the \textit{reciprocal} of $ \alpha $, by denoting it with the symbol $ \alpha^{-1} $. Gibbs resisted this temptation, while Heaviside did not. Indeed, in 1885, by using Clarendon's notation, he had introduced the same symbol by following the quaternion approach, where the reciprocal of a quaternion $ q $ is well defined and denoted by $ q^{-1} $ \cite{Tait-1st}. The reciprocal of a quaternion has the same real part, and the opposite imaginary part, while its norm is the reciprocal of the norm of $ q $. Heaviside's definition of the reciprocal of a vector can be obtained by identifying the three-dimensional vectors with the quaternions that have no scalar part. Indeed, Heaviside defined $ \alpha^{-1} $ as the vector that ``has the same direction of $ \alpha $; its magnitude is the reciprocal of that of $ \alpha $'' (\cite{ELP2}, 5). As we argue in the following, the use of the adjective ``reciprocal'' in the context of vector algebra became one of the contrasting points between Heaviside and Gibbs which originated from Heaviside's analysis of Gibbs's pamphlet. 

Heaviside's criticism appeared on November 18, in 1892, in a paper published in \textit{The Electrician}: `Professor Gibbs calls the vectors  $ \alpha'$,  $ \beta'$,  $\gamma' $ the reciprocals of  $ \alpha$,  $ \beta$,  $\gamma $. They have some of the properties of reciprocals. Thus,
\begin{equation}\label{def-reciprocal}
\alpha\boldsymbol{\cdot}\alpha' = 1\, ,\qquad \beta\boldsymbol{\cdot}\beta'=1\, ,\qquad\gamma\boldsymbol{\cdot}\gamma' = 1\, .
\end{equation}  
But it seems to me that the use of the word reciprocal in this manner is open to objection. It is in conflict with the obvious meaning of the reciprocal $ \alpha^{-1} $ of a vector $ \alpha $ [...]' (\cite{EM1}, 295). Then, he emphasised again: `It will also be observed that Gibbs's reciprocal of a vector depends not upon the vector alone, but upon two vectors as well. It would seem desirable, therefore, to choose some other name than reciprocal. I have provisionally used the word ``complementary'' in the above, to avoid confusion with the more natural use of ``reciprocal''.' (\cite{EM1}, 295). In order to see how Heaviside's point of view changed, we present the evolution of his approach chronologically, following his published papers.
\begin{itemize}
\item[\textbf{1885}] In June, Heaviside introduced the symbol $ \alpha^{-1} $ (\cite{ELP2}, 5). Here, he did not give any special name to the new symbol.
\item[\textbf{1888}] Heaviside received Gibbs's booklet.
\item[\textbf{1891}] \textbf{(April)} Gibbs replied to Tait's statements contained in the preface of the third edition of his \textit{Quaternions}, with a note published on \textit{Nature}.
\item[\textbf{1891}] \textbf{(November, 13)} Heaviside started to publish a series of papers in \textit{The Electrician}, which would reprint subsequently in \textit{Electromagnetic Theory}, which form the first complete published review on vector algebra. Also, Heaviside spoke in favour of Gibbs by commenting the Gibbs-vs-Tait debate. (\cite{EM2}, 137).
\item[\textbf{1891}] \textbf{(December, 18)} By reviewing some concepts introduced in 1885, Heaviside defined $ \alpha^{-1} $ as in 1885: `We define the reciprocal of a vector $ \alpha $ to be a vector having the same direction as $ \alpha $ and whose tensor is the reciprocal of that of  $ \alpha $' (\cite{EM1}, 155). He would adopt the same definition in the \textit{Electrical Papers} on page 530.
\item[\textbf{1892}] \textbf{(January, 1)} Heaviside introduced a set of ``auxiliary vectors'' associated to a set of three non coplanar independent vectors,  which he would recognise as Gibbs's reciprocal set.
\item[\textbf{1892}] \textbf{(June, 12)} Heaviside used explicitly Gibbs's term ``reciprocal set'' (\cite{ELP2}, 22) and introduced the reciprocal dyadic, which we shall discuss in the next section.
\item[\textbf{1892}] \textbf{(November, 18)} Heaviside used for the first time the term ``complementary'' instead of Gibbs's term ``reciprocal''. He recognised that he had implicitly used the set ``complementary to another set of vectors'' on January 1 (\cite{EM1}, 294) and in the following page he criticized Gibbs's choice (quotation cited).
\end{itemize} 

From the above chronology, it can be inferred that Heaviside adopted Gibbs's terminology first, then, following quaternionic tradition, he defined the reciprocal of a single vector and realised that this concept is in conflict with the reciprocal system introduced by Gibbs. Indeed, neither vector belonging to Gibbs's reciprocal system would necessarily satisfy Heaviside's definition of $ \alpha^{-1} $. Therefore, Heaviside criticized Gibbs's choice and introduced a new terminology.

\begin{figure}[ht!]
	\centering
	\includegraphics[width=120mm]{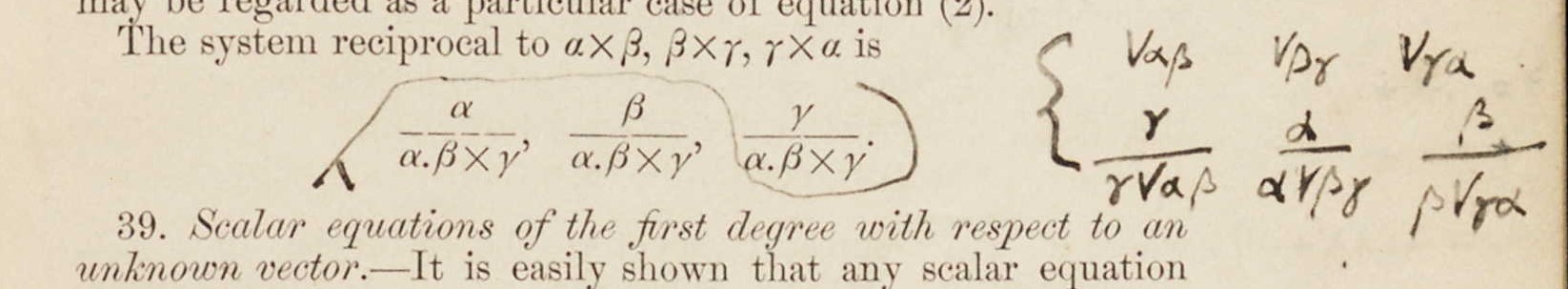}
	\caption{Heaviside's change of order (p. 11)}
	\label{reciprocal3}
\end{figure}

As a consequence of the introduction of the reciprocal of a vector, Heaviside disagreed with Gibbs also in respect to the following fact on reciprocal systems of vectors.
On page 11, given three vectors $ \alpha $, $ \beta $ and $ \gamma $, Gibbs introduced the reciprocal system of the vectors $ \alpha\times\beta $, $ \beta\times\gamma $ and $ \gamma\times\alpha $. As showed in Fig. \ref{reciprocal3}, Heaviside criticised the order proposed by Gibbs. Heaviside's criticism can be interpreted on the basis of the previous annotation. Indeed, Heaviside interpreted Gibbs's text as suggesting that the vector $\displaystyle \frac{\alpha}{\alpha\boldsymbol{\cdot} (\beta\times\gamma)} $ should correspond to the reciprocal vector of $ \alpha\times\beta $ and that the vector  $\displaystyle \frac{\beta}{\alpha\boldsymbol{\cdot} (\beta\times\gamma)} $ should correspond to the reciprocal vector of $ \beta\times\gamma $. Hence, from Heaviside's point of view, this seemed to be an incorrect statement, because e.g. $  \displaystyle \left( \alpha\times\beta\right) \boldsymbol{\cdot}\left(  \frac{\alpha}{\alpha\boldsymbol{\cdot} (\beta\times\gamma)}\right) \neq 1 $.  Therefore, in his annotation, Heaviside changed the order of the vectors for the reciprocal basis. Indeed, by using the cyclic property of the mixed product\footnote{The cyclic property reads: $ \alpha\boldsymbol{\cdot}\left(\beta\times\gamma \right) = \gamma\boldsymbol{\cdot}\left(\alpha\times\beta \right) = \beta\boldsymbol{\cdot}\left(\gamma\times\alpha \right)  $.} $  \displaystyle \left( \alpha\times\beta\right) \boldsymbol{\cdot}\left(  \frac{\gamma}{\alpha\boldsymbol{\cdot} (\beta\times\gamma)}\right) =\left( \beta\times\gamma\right) \boldsymbol{\cdot}\left(  \frac{\alpha}{\alpha\boldsymbol{\cdot} (\beta\times\gamma)}\right) = \left( \gamma\times\alpha
\right) \boldsymbol{\cdot}\left(  \frac{\beta}{\alpha\boldsymbol{\cdot} (\beta\times\gamma)}\right) =  1 $. From Heaviside's point of view, the three vectors
\begin{equation}
	\frac{\gamma}{\alpha\boldsymbol{\cdot} (\beta\times\gamma)}\; ,\quad \frac{\alpha}{\alpha\boldsymbol{\cdot} (\beta\times\gamma)}\; ,\quad\frac{\beta}{\alpha\boldsymbol{\cdot} (\beta\times\gamma)}
\end{equation}
are the reciprocal vectors of $ \alpha\times\beta $, $ \beta\times\gamma $ and $ \gamma\times\alpha $ \textit{respectively}.
But as already said, Gibbs never defined the reciprocal of a single vector. Therefore, from Gibbs's point of view, the order was not important.

The last annotation of this section would regard the extension of the above concepts, i.e. the reciprocal of a dyadic, i.e. a linear operator.
The annotation can be found at the bottom of page 11, where Gibbs discussed how to find the solution of a vector equation.
\begin{figure}[ht!]
	\centering
	\includegraphics[width=90mm]{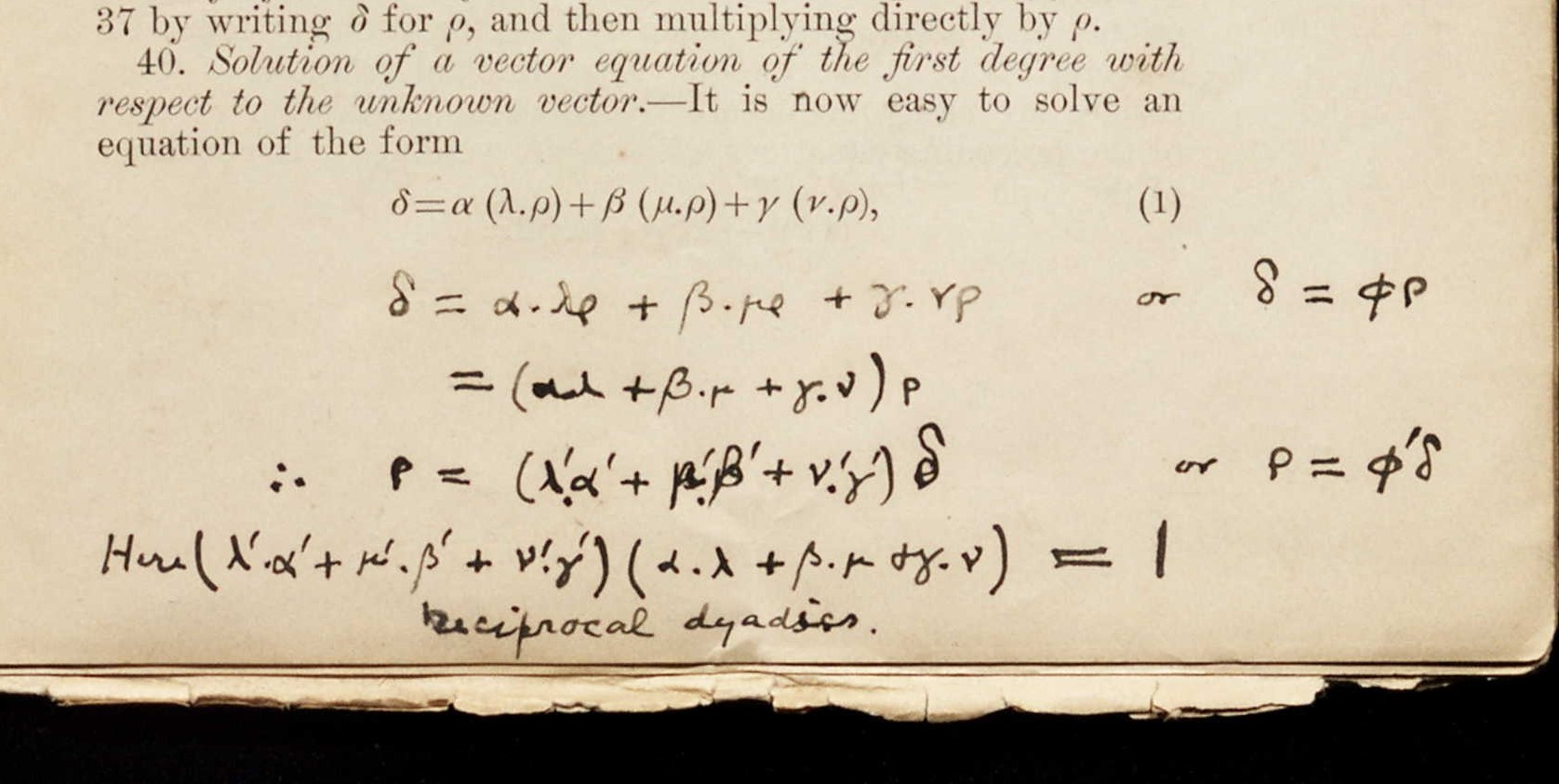}
	\caption{In the first line, Heaviside translated Gibbs's eq. (1). In the second line, he rewrote it by making evident the role of the operator $ \phi $ and in the third line, he inferred that the solution of the vector equation can be obtained by inverting the operator. In the last two lines, Heaviside recognised the concept of the \textit{reciprocal dyadic} (p. 11)}
	\label{vector-eq}
\end{figure}
Heaviside recognised that the problem can be formally solved by inverting the linear operator involved, with the help of reciprocal dyadics, which would generalise the above concept, i.e. the reciprocal of a vector, see Fig. \ref{vector-eq}. 

This annotation played an important role in dating the annotations. Heaviside had introduced the concept of linear operators in his work in July 1883 in the context of the generalised Ohm's law (\cite{ELP1}, 286), but in this annotation, he used the term dyadic, a term coined by Gibbs two chapters later. As already said, Heaviside must have read all Gibbs's pamphlet when he made this annotation, because he had never used this term in his published works, before 1892. Heaviside introduced the term dyadic in his \textit{Electromagnetic Theory} (\cite{EM1}, 263). In order to discuss abstract properties of linear operators, Heaviside introduced Gibbs's dyadic representation (\cite{EM1}, 285). In this context, by discussing the inversion of a linear operator, Heaviside emphasised that  `the simple way [to invert a linear operator] is in term of dyads' (\cite{EM1}, 293). From our point of view, this comment can be traced back to this annotation and it is evidence of Gibbs's influence on Heaviside's work, implied by the analysis of Gibbs's booklet.

\section{Heaviside's annotations on linear operators}\label{lin-operator-section}
In chapters one and two, given three vectors $ \alpha $, $ \lambda $ and $ \rho $, Gibbs considered  expressions like $ \alpha\left( \lambda \boldsymbol{\cdot}\rho\right) $, and he always wrote them without omitting the parentheses. At the beginning of chapter three, he introduced linear vector functions, whose generic expression reads `$ \alpha\lambda \boldsymbol{\cdot}\rho + \beta\mu \boldsymbol{\cdot}\rho + \text{etc.} $' (\cite{Gibbs-pamphlet}, 40), and he omitted the parentheses. Then, in order to introduce linear operators, he wrote `$ \left\lbrace  \alpha\lambda + \beta\mu + \text{etc.} \right\rbrace \boldsymbol{\cdot}\rho $' (\cite{Gibbs-pamphlet}, 40), where the expression in parentheses is a formal definition of the linear operator associated to the six vectors involved. The operator is implicitly acting on the right and Gibbs emphasised that the above expression differs in general from `$ \rho \boldsymbol{\cdot}\left\lbrace  \alpha\lambda + \beta\mu + \text{etc.} \right\rbrace  $' (\cite{Gibbs-pamphlet}, 40), where the operator acts on the left, but both expressions represent vector functions. Here, Gibbs named \textit{dyad} the new object $ \alpha\lambda $ and called dyadic a linear combination of dyads. Gibbs defined implicitly the symbol $ \alpha\lambda $ as the linear transformation drawing a correspondence from a generic vector $ \rho $ to  $ \alpha (\lambda\cdot\rho ) $, which is the modern definition of the tensor product $ \alpha\otimes\lambda $ (\cite{I-Shih}, 238). In Gibbs's language, the dyad represented by $ \alpha\lambda $ should be understood as the indeterminate product between $ \alpha $ and $ \lambda $. For dyadics, Gibbs introduced capital Greek letters. Hence, a linear vector function acting on the right on a generic vector $ \rho $, i.e. `$ \left\lbrace  \alpha\lambda + \beta\mu + \text{etc.} \right\rbrace \boldsymbol{\cdot}\rho $' (\cite{Gibbs-pamphlet}, 40), is represented in short by `$ \Phi \boldsymbol{\cdot}\rho $' (\cite{Gibbs-pamphlet}, 41), which using index notation reads: $ \Phi_{ij}\rho_j $. On page 40, given a generic vector function $ \tau $, Gibbs defined $ \alpha $, $ \beta $ ad $ \gamma $ as `the values of $ \tau $'. With this generic phrase, Gibbs intended that, given a canonical basis of the three-dimensional space $ \vec{i} $, $ \vec{j} $ and $ \vec{k} $, and by calling $ \Phi $ the operator which maps the generic vector $ \rho $ into the generic vector $ \tau $, we can define either $ \alpha = \Phi\cdot\vec{i} $; $ \beta = \Phi\cdot \vec{j} $ and $ \gamma = \Phi\cdot \vec{k} $, or $ \alpha = \vec{i}\cdot \Phi $; $ \beta = \vec{j}\cdot \Phi $ and $ \gamma = \vec{k}\cdot\Phi $. Indeed, because the operator can act either on the right or on the left, Gibbs emphasised that a generic linear function may be expressed either by a dyadic as a pre-factor, namely (\cite{Gibbs-pamphlet}, 41):
\begin{equation}\label{tau1}
\tau =\left\lbrace  \alpha \vec{i} + \beta \vec{j} + \gamma \vec{k}\right\rbrace \boldsymbol{\cdot}\rho\, ;
\end{equation}
or by a dyadic as a post-factor, namely (\cite{Gibbs-pamphlet}, 41):
\begin{equation}\label{tau2}
\tau =\rho \boldsymbol{\cdot}\left\lbrace  \vec{i}\alpha + \vec{j}\beta  + \vec{k}\gamma \right\rbrace \;\, . 
\end{equation}
Even if Gibbs used the same letter, equations [\ref{tau1}] and [\ref{tau2}] do not refer to the same linear function, unless the matrix representing the linear is symmetrical, as the author would state in a subsequent chapter. 

In the following paragraph, we will show how Heaviside's criticism toward the indeterminate product originated during his studying of Gibbs's booklet. Then, we shall present a new mathematical object introduced by Heaviside as a development of Gibbs's ideas.

\subsection{On the ``indeterminate'' product}\label{indeterminate}
\begin{figure}[ht!]
	\centering
	\includegraphics[width=100mm]{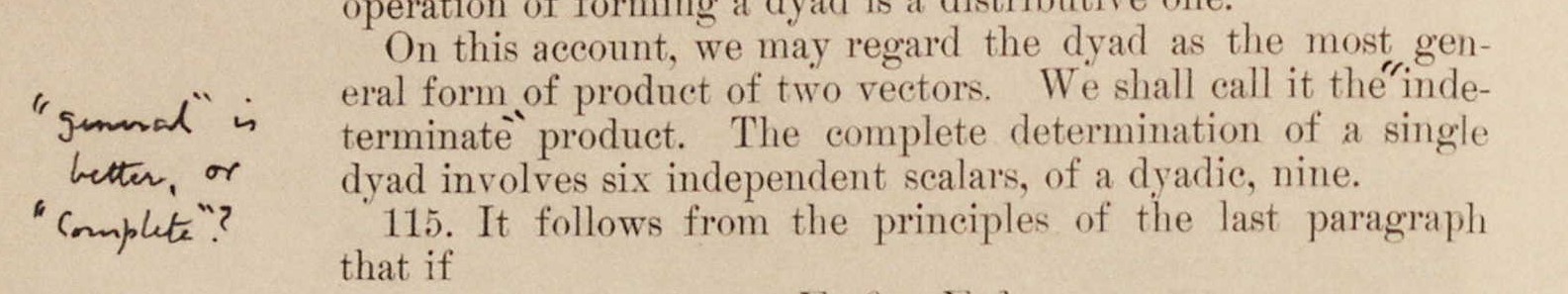}
	\caption{Heaviside's critique of Gibbs's \textit{indeterminate} product (p. 42)}
	\label{indeterminate-fig}
\end{figure}

Heaviside's annotation on page 42 is a critique of the name that Gibbs used for the tensor product. Gibbs wrote: `we may regard the dyad as the most general form of the product of two vectors. We shall call it the \textit{indeterminate product}.' [emphasis added] (\cite{Gibbs-pamphlet-2},42). Heaviside did not like the adjective indeterminate. He inserted quotation marks around it and wrote on the margin: ` ``general'' is better, or ``complete''? ' (\cite{Gibbs-pamphlet}, 42), see Fig. \ref{indeterminate-fig}.
Heaviside never defined such a product or gave it a particular name, neither in \textit{Electrical Papers} nor in \textit{Electromagnetic Theory}. As we said, Heaviside used the period in order to `keep the proper symbols connected' (\cite{EM1}, 179) and this was enough for him. 

From Gibbs's point of view, the name of the product represented more than a mere label. Indeed, it was connected with the interpretation of the product itself. Gibbs had clarified his point of view in a published work in 1886 (\cite{Gibbs-scientific}, 91), written in order to call attention on the influence of Hermann G. Grassmann on the theory of multiple algebras\footnote{The algebra of complex numbers is an example of \textit{double} algebra. Multiple algebras are generalisations of this mathematical structure.} \cite{Grassmann}. By discussing the concept of the product, Gibbs compared the approaches of Hamilton, Augustus De Morgan and Charles S. Peirce on one side, and that of Grassmann on the other side. While the formers spoke of the product of two \textit{multiple quantities}, i.e. elements of a multiple algebra, `as if only \textit{one} product could exist, at least in the same algebra.'[emphasis added] (\cite{Gibbs-scientific}, 105), the latter seemed to look at the concept of product in a more abstract way. Indeed, Grassmann had introduced in 1844 the concept of \textit{open} product, which corresponded to the indeterminate product, from Gibbs's point of view. Grassmann defined it as \textit{open} product, because the result of the operation is a new object as it happens for multiple applications of operators (\cite{Gibbs-scientific}, 104). Gibbs recognized that he followed independently Grassmann's approach, when he addressed the question, as the German mathematician did, `what products, i.e. what distributive functions of the multiple quantities, are most important?' (\cite{Gibbs-scientific}, 105). Gibbs's answer follows. `Let us return to the indeterminate product, which I am inclined to regard as the most important of all' (\cite{Gibbs-scientific}, 109). Why is it the most important and hence, as he wrote in his booklet, the most general? First, Gibbs called attention on the fact that Grassmann introduced also the scalar, which he called internal, product and the external product, which is, as we know today, a generalisation of the vector product. Then, he emphasised that both products can be defined by using the tensor product. Therefore, we can infer that for this reason Gibbs considered the tensor product as the most general one. Before proceeding, it is worth noticing that this paper is important also because Gibbs called attention on the connection between multiple algebra and matrices. 

Gibbs repeated his arguments in a letter to Victor Schlegel in 1888. Regarding his paper on multiple algebras, Gibbs asserted again that he wrote it `to call attention to the fundamental importance of Grassman's work in this field, \& lastly, to express my own ideas on the subject' (\cite{Gibbs-biblio}, 107). Then he underlined: `My dyadic are not \textit{algebraic} product, but the most general product mentioned by Grassmann, those having \textit{no special} law, say indeterminate products.' [emphasis added] (\cite{Gibbs-biblio}, 107). Hence, the product's name is also connected with the fact that commutative and associative laws can be viewed as special characters for a product.

Gibbs's pupil Wilson would use the same adjective in his book based on Gibbs's booklet. Wilson motivated the choice of the name as follows\footnote{Wilson used Gibbs's notation for the symbols denoting the products, but used Latin letters in bold form for the vectors.}. `The reason for the term indeterminate is this. The two products $ \vec{a}\boldsymbol{\cdot}\vec{b} $ and $ \vec{a}\times\vec{b} $ have definite meanings. One is a certain scalar, the other a certain vector. On the other hand, the product $ \vec{a}\,\vec{b} $ is neither vector nor scalar -- it is purely symbolic and acquires a determinate physical meaning only when used as an operator. The product $ \vec{a}\vec{b} $ does not obey the commutative law. It does, however, obey the distributive law [...] and the associative law as far as scalar multiplication is concerned [...].' (\cite{Wilson}, 272).

On March 21, 1902, Heaviside wrote a review of Wilson's work where his critique appeared. `He [Gibbs] calls $ \phi $ a dyadic, and $ \vec{a}\,\vec{l} $ is called a dyad. This brings me to the somewhat important questions of notation, and what should be called a product. [...] $ \vec{a}\,\vec{l} $ itself without any mark between [...], he [Gibbs] calls the indeterminate product and says it is the most general product. I have great respect for Professor Gibbs, and I have carefully read what Dr Wilson says in justification of regarding the dyad as a product. But \textit{I have failed to see that it is a product at all}. The arguments seem very strained, and I think this part of Gibbs's dyadical work will be difficult for students. In what I write $ \vec{a}.\vec{l} $, the dot is a separator; $ \vec{a}$ and $ \vec{l} $ do not unite in any way. With a vector operand $ \vec{r}$ we get $ \vec{a}.\vec{l}\vec{r} $, [...]. That is plain enough, but I do not see any good reason for considering the operator $ \vec{a}.\vec{l} $ to be the general product, in whatever notation it may be written.'[emphasis added] (\cite{EM3}, 141). Hence, Heaviside criticised not only the name of the product but also the essence of the concept itself. In spite of this, Gibbs's term would survive for at least forty years, see for example (\cite{Hitchcock}, 165). 

\subsection{Heaviside's new triadic}\label{triadic-section}
In this section, we shall use both Gibbs and Heaviside's notation. Heaviside's formulas will be preceded by the symbol [\textbf{H}]. It is worth remembering that Heaviside used the period as a mere separator and that he used capital letter $ V $ to denote vector product, while for the scalar product he introduced no symbols. In the annotation we present in this section, Heaviside would introduce a new object by merging Gibbs's approach and the quaternionist one.  

\begin{figure}[ht!]
	\centering
	\includegraphics[width=110mm]{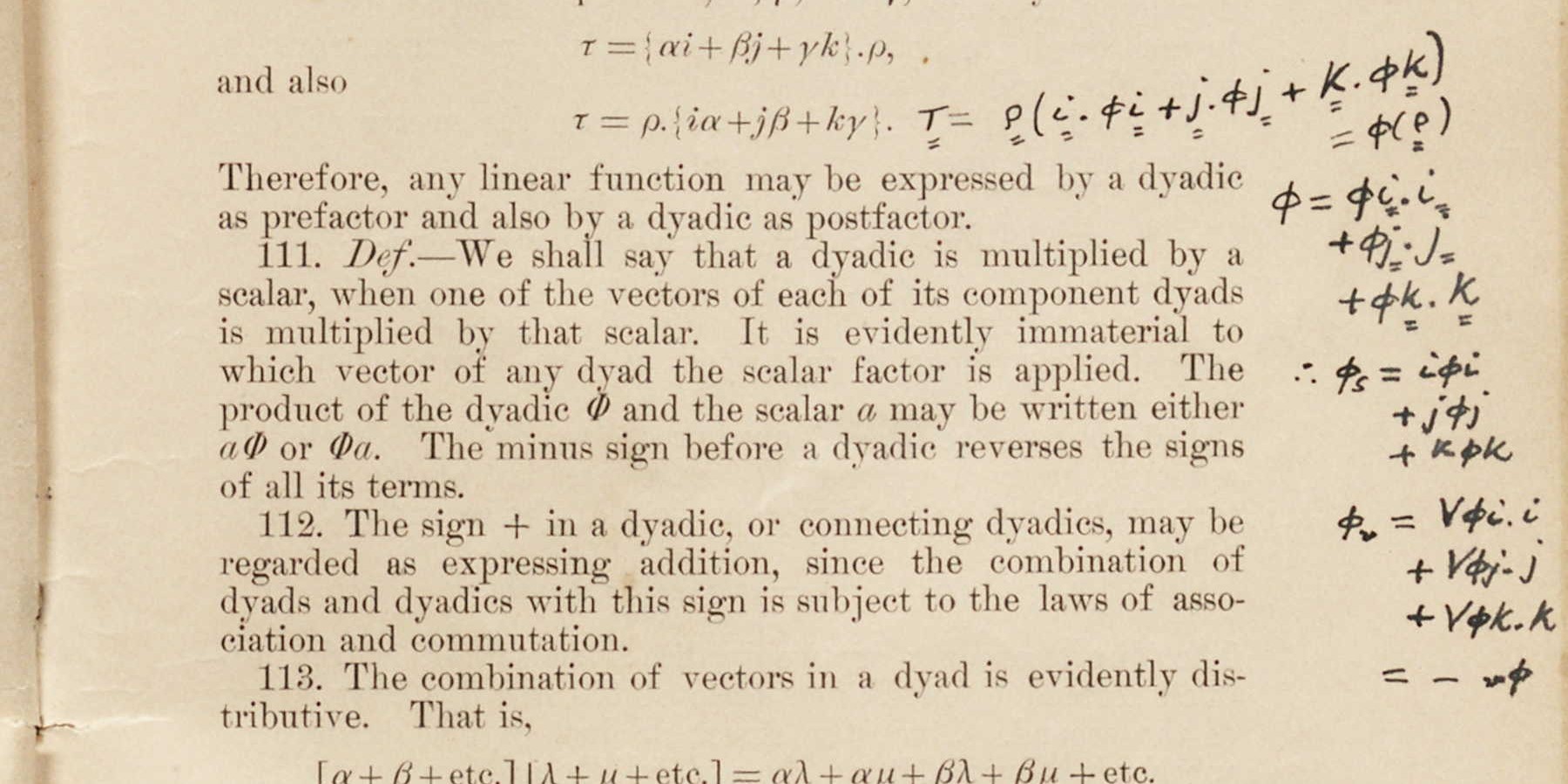}
	\caption{\textbf{First and second line:} Heaviside's translation. \textbf{Third, fourth and fifth line:} abstract representation of the operator $ \Phi $. \textbf{Sixth, seventh and eighth line:} trace of the operator $ \Phi $. \textbf{Last four lines:} Heaviside's new triadic (p. 41)}
	\label{new-triadic}
\end{figure}

In the first line of Fig. \ref{new-triadic}, Heaviside translated the vector function $ \tau $, equation [\ref{tau2}], and inserted [\textbf{H}] $ \alpha=\Phi \vec{i} $, and so on. Hence, he explicitly identified the vector function as the evaluation of the operator $ \Phi $ on the generic vector by writing `$ = \Phi (\rho) $' (second line of Fig. \ref{new-triadic}). Heaviside marked with doubled underlines all the vector quantities.
It is worth noting that he should have used [\textbf{H}] $ \tilde{\alpha}=\vec{i}\Phi  $, as he correctly did in the third and following lines. After having formally rewritten the operator as [\textbf{H}] $ \Phi = \Phi \vec{i}.\vec{i} + \Phi \vec{j}.\vec{j} + \Phi \vec{k}.\vec{k} $ in the third, fourth and fifth lines, he inferred that two new different objects could be defined: a scalar one and a ``vector'' one. Maybe Heaviside tried to separate the original dyadic into a scalar and a vector part, as quaternionists did for quaternions. The scalar $ \Phi_s $ will be defined also by Gibbs on page 42 and it corresponds to the trace of the linear operator. It can be obtained by substituting the tensor product with the scalar product (see also \cite{Wilson}, 275). Heaviside's vector quantity $ \Phi_v  $, instead, never appears in Gibbs's booklet. Indeed, it should not be confused with $ \Phi_{\times} $, introduced by Gibbs on page 42. Gibbs's $ \Phi_{\times} $ is a vector quantity and can be obtained by substituting the tensor product with the vector product in the definition of $ \Phi $. Heaviside's $ \Phi_v  $ is a triadic because he substituted the scalar product with the vector product between the dyadic $ \Phi  $ and the vector to which it is applied. By using the index notation, the difference emerges more clearly as follows.
\begin{eqnarray}
& \text{Gibbs:} &\quad \left(\,  \Phi_{\times}\,\right)_{k}\;\, =\; \Phi_{ij}\epsilon_{ijk}\label{phi-cross}\\
& \text{Heaviside:} & \quad\left(\, \Phi_v\,\right)_{ijk}=\; \Phi_{im}\epsilon_{mjk}\label{phi-vec}\; .
\end{eqnarray}

Heaviside had already introduced the analogous of Gibbs's $ \Phi_{\times} $ before 1888. Indeed, on January 15, 1886 \textit{The Electrician} published a paper where Heaviside introduced it in order to describe `the torque per unit volume', when `there is a rotational force arising from stress' acting on a body (\cite{ELP1}, 544). The new triadic $ \Phi_v  $ instead, as far as we are aware, was never defined by Heaviside in his published work. On the other side, he used it implicitly in order to define the cross product between dyadics and vectors, e.g. $ \Phi\times\rho $ (\cite{EM1}, 295, equations (145) and (146)). The same definition appears in Gibbs's booklet (\cite{Gibbs-pamphlet}, 43). Heaviside was aware of the triadic character of eq. [\ref{phi-vec}], because in his \textit{Electromagnetic Theory} he observed that `whereas the direct product $ \Phi\vec{r} $ of the dyadic $ \Phi $ and a vector $ \vec{r} $ is a vector; on the other hand, the skew products $ V\Phi\vec{r} $ and $ V\vec{r}\Phi $ are themselves dyadic.' (\cite{EM1}, 295). Heaviside's $ V\Phi\vec{r} $ can be obtained by applying the triadic  $ \Phi_v  $, eq. [\ref{phi-vec}], on the vector $ \vec{r} $. In addition, Heaviside emphasised that the scalar product between the vectors $ \vec{s} $ and $ V\Phi\vec{r} $, i.e. [\textbf{H}] $ \vec{s}V\Phi\vec{r} $, is a vector (\cite{EM1}, 295). Indeed, using index notation, it corresponds to $ s_i\Phi_{im}\epsilon_{mjk}r_j $. In \textit{Electromagnetic Theory}, after these statements, the author did not further investigate this concept, because, as he himself wrote, it was beyond the scope of his treatment. 

Before proceeding, it should be mentioned that the last line of Heaviside's annotation is not clear. Indeed, it seems that Heaviside wanted to define an operator obtained with the skew product on the left side, i.e. $ _v\Phi $, and that the following operator identity should hold [\textbf{H}]: $ \Phi_v = -_v\Phi $. In \textit{Electromagnetic Theory} no such property is mentioned. Let us translate Heaviside's $ V\Phi\vec{r} $ and $ V\vec{r}\Phi $ in component notation, namely: 
\begin{eqnarray}
\left( \,V\Phi\vec{r}\,\right)_{ik}  &=&\, \Phi_{im}\epsilon_{mjk}r_j\\
\left(\, V\vec{r} \Phi\,\right)_{ik}  &=&\, r_j\Phi_{mi}\epsilon_{jmk} = -\Phi_{mi}\epsilon_{mjk}r_j =-\left( \Phi_C\right)_{im}\epsilon_{mjk}r_j  \, ,\label{identity}
\end{eqnarray} 
where we used the antisymmetric property of the Levi-Civita symbol and the definition of Gibbs's conjugated dyadic, namely $\displaystyle  \left(\, \Phi_C \,\right)_{im} \overset{def}{=} \Phi_{mi} $. From eq. [\ref{identity}] it follows that $ V\Phi\vec{r} =- V\vec{r} \Phi$ holds if and only if $ \Phi_{im}=\Phi_{mi} $, or, in other words, for symmetric operators. The general property following from eq. [\ref{identity}] is $ V\Phi\vec{r} = - V\vec{r} \Phi_C$. It is worth noting that in \textit{Electromagnetic Theory} (\cite{EM1}, 286) as well as in the annotations made on the booklet, Heaviside used primed letters to indicate conjugate operators, hence, maybe, a prime sign is simply missing for unknown reasons.

\subsection{On the \textit{curl} of a tensor}
On page 66 of Gibbs's booklet, the author discussed how to apply the $ \nabla $ operator to dyadics. In paragraph 161, he defined the action of $ \nabla\times $ and $ \nabla\boldsymbol{\cdot} $ on a dyadic $ \Phi $ as follows (\cite{Gibbs-pamphlet-2}, 66):
\begin{eqnarray}
\nabla\times\Phi &\overset{def}{=}& \vec{i}\times \frac{d\Phi}{dx}+\vec{j}\times\frac{d\Phi}{dy}+\vec{k}\times\frac{d\Phi}{dz}\label{curl-tens}\\
\nabla\boldsymbol{\cdot}\Phi&\overset{def}{=}& \vec{i}\boldsymbol{\cdot} \frac{d\Phi}{dx}+\vec{j}\boldsymbol{\cdot}\frac{d\Phi}{dy}+\vec{k}\boldsymbol{\cdot}\frac{d\Phi}{dz}\label{div-tens}\; ,
\end{eqnarray}
where $ i $, $ j $ and $ k $ are now the canonical basis of the three-dimensional space. Heaviside already knew equation [\ref{div-tens}] because he had introduced it in a previous work. Instead, as far as we know, he had never considered the analogous of equation [\ref{curl-tens}], which defines the \textit{curl} of a tensor. This mathematical object attracted Heaviside's attention. In the following, first we shall contextualise and analyse Heaviside's annotation. Second, we shall make some comments on the \textit{curl} of a tensor in the context of Heaviside's approach to electromagnetic theory. Then, we shall discuss the emergence of this mathematical concept in the context of the theory of elasticity. 

In order to understand Heaviside's interest and his annotation, we go back to 1886. On January 15, the \textit{Electrician} published a paper by Heaviside on mechanical forces and stresses. Heaviside's aim was to discuss what he called `Maxwellian Stresses' (\cite{ELP1} p. 542), i.e. an electromagnetic analogue of the tensor of stresses, in the context  of continuum mechanics. First, Heaviside introduced what he called `the simple stresses (pressures or tensions)'  (\cite{ELP1} p. 543). He did not use dyadics, but he defined three vectors, namely $ \vec{P}_1 $, $ \vec{P}_2 $ and $ \vec{P}_3 $, as it is usual nowadays also in continuum mechanics, which represent `the vector stresses per unit area on planes whose normals are $ \vec{x} $, $ \vec{y} $, $ \vec{z} $ respectively.' (\cite{ELP1} p. 543), i.e. pressures. Heaviside specified: `These are the forces exerted by the matter on the positive side on that on the negative side of the three planes, and, being forces, are vectors.'  (\cite{ELP1} p. 543). Unlike Gibbs, Heaviside used $ \vec{i} $, $ \vec{j} $ and $ \vec{k} $ for representing the \textit{canonical basis} of the three-dimensional space. Having named $ P_{11} $, $ P_{12} $ and $ P_{13} $ the three components of $ \vec{P}_1 $, and analogously for the other two, the three vector stresses read:
\begin{eqnarray}\label{H-stress}
	 \vec{P}_1 &=& \vec{i}P_{11}+\vec{j}P_{12}+\vec{k}P_{13}\, ,\nonumber\\ 
	 \vec{P}_2  &=& \vec{i}P_{21}+\vec{j}P_{22}+\vec{k}P_{23 }\, ,\\ 
	 \vec{P}_3  &=& \vec{i}P_{31}+\vec{j}P_{32}+\vec{k}P_{33}\,\nonumber  
\end{eqnarray}
Equations (\ref{H-stress}) formally define a tensor, which is represented by a $ 3\times 3 $ -matrix whose components are $ P_{ij} $. Then, Heaviside specified that in the case of mechanical stresses the matrix is symmetric, i.e. $ P_{ij}= P_{ji} $, and that in this case `the translational force due to stress' (\cite{ELP1} p. 543) can be defined as follows:
\begin{equation}\label{Pot-Force-H}
\vec{i}\, \text{div}\vec{P}_1 + \vec{j}\, \text{div}\vec{P}_2 + \vec{k}\, \text{div}\vec{P}_3\; ,
\end{equation}
which in components notation reads $\displaystyle
F_i=\partial_j P_{ij} $. Heaviside speculated on the possibility that equation [\ref{Pot-Force-H}] could be a particular case of a more general expression. Therefore, he supposed that the correct definition of the translational force should be 
 \begin{equation}\label{Pot-Force}
F_i = \partial_j P_{ji} \; ,
 \end{equation}  
because it coincides with the previous one for symmetrical matrices. Furthermore, he introduced the analogue of Gibbs's $ \Phi_{\times} $, see eq. [\ref{phi-cross}], which corresponded to the `torque per unit volume' (\cite{ELP1} p. 544) and which in components notation reads $ \epsilon_{ijk}P_{ij} $. Heaviside repeated quite a similar presentation in 1891, i.e. after having received and read Gibbs's booklet\footnote{As we said in section \ref{dating}, Heaviside emphasised the importance of Gibbs's results in the context of linear operators.} (\cite{ELP2}, 533). As we said above, equation [\ref{curl-tens}] aroused Heaviside's interest. Indeed, we found two different annotations on this subject: the first on page 66 and the second on the back cover of the booklet itself.

In his first annotation, Heaviside recognised the connection between Gibbs's abstract definition and the theory of continuum mechanics.
\begin{figure}[ht!]
	\centering
	\includegraphics[width=100mm]{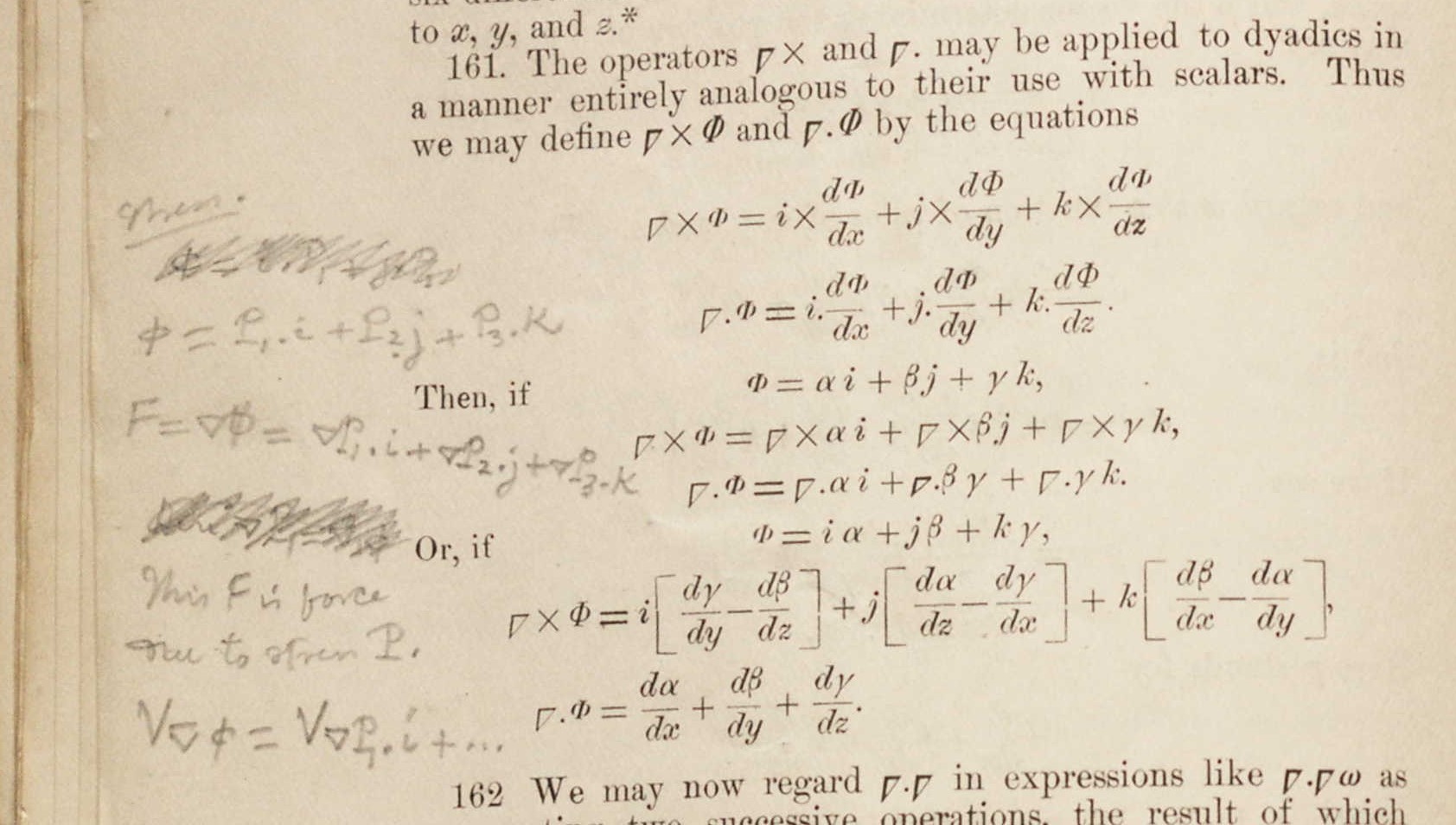}
	\caption{Divergence and \textit{curl} of the stress tensor (p. 66)}
	\label{curl-figure-1}
\end{figure} 
Indeed, as shown in Fig. \ref{curl-figure-1}, Heaviside associated the dyadic $ \Phi $ to the stress tensor $ P_{ij} $ by writing (first two rows of Fig. \ref{curl-figure-1}) [\textbf{H}]: 
\begin{eqnarray}
	&\text{Stress:}&\nonumber\\
	&\Phi & =\quad \vec{P}_1\, .\, \vec{i} + \vec{P}_2\, .\, \vec{j} + \vec{P}_3\, .\, \vec{k}\; ,\label{H-stress-2}
\end{eqnarray}
which is a contracted form for equations [\ref{H-stress}]. Then, he recognised the role of eq. [\ref{div-tens}] by annotating (second, third and fourth row of Fig. \ref{curl-figure-1}) [\textbf{H}]:
\begin{eqnarray}
	& F & =\;\; \nabla\Phi\; =\;
	 \nabla \vec{P}_1\, .\, \vec{i} +\nabla \vec{P}_2\, .\, \vec{j} +\nabla \vec{P}_3\, .\, \vec{k}\label{Pot-Force-2}\\
	&\text{This}\; F& \text{is force}\nonumber\\
	&\text{due to}&\text{stress}\; P
\end{eqnarray}
In the last row, Heaviside simply translated Gibbs's eq. [\ref{curl-tens}] [\textbf{H}]:
\begin{equation}
V\nabla\Phi = V\nabla \vec{P}_1\, .\,\vec{i} + \dots\label{curl-tens-H}
\end{equation}

The second annotation is on the back cover of the booklet, where Heaviside continued to consider the context of continuum mechanics. The back cover is the most damaged part of the pamphlet and it is reported on appendix \ref{app1}, where we presented a complete transcription of the page. Here we shall make our comments by rewriting it step by step.

As already said, Heaviside was interested in Gibbs's eq. [\ref{curl-tens}], maybe because he did not realise if it had some physical meaning. Indeed, on the back cover, firstly he wrote [\textbf{H}]:
\begin{eqnarray}
\nonumber\\
\text{Stress.} &\text{What}&\text{is\, [the physical]\, meaning\, of}\,\; curl\, \phi\, ?\nonumber\\
\nonumber\\
&\phi & = \vec{P}_1\, .\,\vec{i} + \vec{P}_2\, .\,\vec{j} + \vec{P}_3\, .\,\vec{k} \nonumber \\
&\nabla\phi & = F = \nabla \vec{P}_1\, .\,\vec{i} +\nabla \vec{P}_2\, .\,\vec{j} +\nabla  \vec{P}_3\, .\,\vec{k} \nonumber\\
& V\nabla\phi & = G =  V\nabla \vec{P}_1\, .\,\vec{i} +V\nabla \vec{P}_2\, .\,\vec{j} +V\nabla  \vec{P}_3\, .\,\vec{k} \label{back-1}\\
\nonumber\\
\phi\,&\text{is stress} &\text{tensor;}\; F\,\text{is force vector. What is}\; G\, ?\nonumber
\end{eqnarray}
where the square-bracketed word cannot be read, because the writing has faded. In the third line of eq. [\ref{back-2}] the former of Heaviside's annotation appears, i.e. eq. [\ref{curl-tens-H}]. This fact establishes a connection between the two annotations. On the rest of the back cover, Heaviside tried to understand what are the components of $ G $, i.e. $ curl\, \phi $ [\textbf{H}]:
\begin{eqnarray}
G &=& V\vec{i}\nabla .\;\phi\;\;\text{etc}\nonumber\\ 
 &\stackrel{\arrowvert}{=}& \vec{i}V\nabla .\;\phi\label{back-2}
\end{eqnarray}
He developed equation [\ref{back-2}] by writing the first column vector explicitly as follows\footnote{All the derivatives must be intended as partial derivatives. The modern symbol for partial derivatives was not yet commonly used at that time.} [\textbf{H}]:
\begin{eqnarray}
V\vec{i}\nabla\, .\,\phi = \left( \vec{k}\frac{d}{dy}-\vec{j}\frac{d}{dz}\right) \phi &=& \left( \vec{k}\frac{d}{dy}-\vec{j}\frac{d}{dz}\right)\left( \vec{P}_1\, .\,\vec{i} + \vec{P}_2\, .\,\vec{j} + \vec{P}_3\, .\,\vec{k} \right) \label{back-3a}\\
&=& \left( -\frac{dP_{12}}{dz}+\frac{dP_{13}}{dy}\right) .\; \vec{i}+\left( -\frac{dP_{22}}{dz}+\frac{dP_{23}}{dy}\right) .\; \vec{j} + \left( -\frac{dP_{32}}{dz}+\frac{dP_{33}}{dy}\right) .\; \vec{k}\label{back-3b}\\
&=& \vec{i}V\nabla \vec{P}_1 .\;\vec{i} + \vec{i}V\nabla \vec{P}_2 .\;\vec{j} + \vec{i}V\nabla \vec{P}_3 .\;\vec{k}\label{back-3c}\\ 
&=& \vec{i}V\nabla .\;\phi\label{back-3d}
\end{eqnarray}
In the first line, eq. [\ref{back-3a}], Heaviside calculated firstly the vector product between the vector $ \vec{i} $ and the operator $\displaystyle
\nabla \overset{def}{=} \vec{i}\frac{d}{dx}+\vec{j}\frac{d}{dy}+\vec{k}\frac{d}{dz}$, and then he inserted the definition of $ \phi $. In the second line, eq. [\ref{back-3b}], he applied the operator $ \displaystyle \left( \vec{k}\frac{d}{dy}-\vec{j}\frac{d}{dz}\right) $ to the dyadics, starting with the closest operator. This means that he applied $  \displaystyle  -\vec{j}\frac{d}{dz} $ to the vector $ \vec{P}_1 $ and then $\displaystyle \vec{k}\frac{d}{dy} $ to the same vector. At the bottom of the page, the explicit calculation of this first part appears:
\begin{equation}
\frac{d\vec{P}_3}{dy} - \frac{d\vec{P}_2}{dz}\,\quad =\quad \vec{i}\left( \frac{dP_{13}}{dy}-\frac{dP_{12}}{dz}\right) +\vec{j}\left( \frac{dP_{32}}{dy}-\frac{dP_{22}}{dz}\right) + \vec{k}\left( \frac{dP_{33}}{dy}-\frac{dP_{23}}{dz}\right) \label{rot-phi}     
\end{equation} 
Indeed, in order to write equation [\ref{rot-phi}], Heaviside interpreted the matrix $ \phi $ as a row of vectors, see eq. [\ref{H-stress}], and calculated the vector product with the operator $ \nabla $. Hence, he used eq. [\ref{rot-phi}] to infer the equivalence between equations [\ref{back-3c}] and [\ref{back-3d}].

Finally, Heaviside wrote in the centre of the page the complete expression for the \textit{curl} of the dyadic $ \phi $, namely
\begin{eqnarray}
G & = & \left[ \vec{i}\left( \frac{dP_{13}}{dy}-\frac{dP_{12}}{dz}\right) +\vec{j}\left( \frac{dP_{11}}{dz}-\frac{dP_{13}}{dx}\right) + \vec{k}\left( \frac{dP_{12}}{dx}-\frac{dP_{11}}{dy}\right) \right] .\; \vec{i} \nonumber\\
& + & \left[ \vec{i}\left( \frac{dP_{23}}{dy}-\frac{dP_{22}}{dz}\right) +\vec{j}\left( \frac{dP_{21}}{dz}-\frac{dP_{23}}{dx}\right) + \vec{k}\left( \frac{dP_{22}}{dx}-\frac{dP_{21}}{dy}\right) \right] .\; \vec{j} \nonumber\\
& + & \left[ \vec{i}\left( \frac{dP_{33}}{dy}-\frac{dP_{32}}{dz}\right) +\vec{j}\left( \frac{dP_{31}}{dz}-\frac{dP_{33}}{dx}\right) + \vec{k}\left( \frac{dP_{32}}{dx}-\frac{dP_{31}}{dy}\right) \right] .\; \vec{k} \label{curl-phi}\; ,
\end{eqnarray}
which reads by using the components notation:
\begin{equation}\label{curl-phi-components}
G_{lm} = \left( curl\,\phi\right)_{lm} =  \epsilon_{ijl}\partial_i\phi_{mj} \; .
\end{equation}

The \textit{curl} of a tensor is a mathematical object which has different applications in Physics. During the 1890s, i.e. when Heaviside was studying Gibbs's booklet, the applicability of this concept had not emerged explicitly yet. In the following part, we shall discuss briefly the emergence of this concept in the three frameworks we considered: the context of electromagnetic theory in Heaviside's published work, the context of the theory of elasticity, where Heaviside asked himself the meaning for the first time, and finally the context of abstract vector calculus in Gibbs's pamphlet.

As far as we know, neither volumes of the \textit{Electrical Papers} nor of the \textit{Electromagnetic Theory} contain an explicit discussion of this mathematical tool, i.e. the \textit{curl} of a tensor, and Heaviside never addressed explicitly the question of its physical meaning. In spite of this, some of the formulas used by Heaviside seem to involve implicitly this concept in the electromagnetic context. Since July 1883, Heaviside had been considering the existence of electrical `eolotropy' (\cite{ELP1}, 286), i.e. the phenomenon whereby the electric conductivity of a body depends on the direction in which it is measured. By generalising Ohm's law, Heaviside introduced the concept of linear operators in order to describe the eolotropy phenomenon. On February 21, 1885, a paper where Heaviside investigated the role of eolotropy in Maxwell's equations appeared in the \textit{Electrician}. In this context, the British scientist wrote the following expression (\cite{ELP1}, 541, equation (32)):
\begin{equation}\label{eq-curl-E}
 	curl\,\mu^{-1}\, curl\,\vec{E} + c\ddot{\vec{E}} = 0 \; ,
\end{equation}
where $ \vec{E} $ is the electric field, the dots represent the partial second time-derivative, and $ c $ and $ \mu^{-1} $ are linear operators, which represented the permittivity and the inverse of the permeability of a medium respectively. Equation [\ref{eq-curl-E}] can be read in two different ways. It can be claimed, as we emphasised, that it contains the \textit{curl} of the operator $ \mu^{-1} $ applied to the \textit{curl} of the electric field. Conversely, it could be interpreted as the \textit{curl} of a vector $ \vec{v} $, which is obtained by applying the operator $ \mu^{-1} $ to the \textit{curl} of the electric field. Indeed, the two interpretations are equivalent. Expressions like eq. [\ref{eq-curl-E}] appeared again in Heaviside's work. But if we consider his investigation on the origin of double refraction in the context of an elastic ether released at the end of 1893 in the \textit{Proceedings of the Royal Society of London} (\cite{EM2}, 518), an interesting fact emerges. After having received Gibbs's booklet, when considering linear operators, Heaviside not only quoted Gibbs's term dyadic, but also adopted explicitly Gibbs's dyadical form. `In Maxwell's electromagnetic theory, the two properties are those connecting the electric force with the displacement, and the magnetic force with the induction, say the permittivity and the inductivity, or $ c $ and $ \mu $. These are, in the simplest case, constants corresponding to isotropy. The existence of eolotropy as regards either of them will cause double refraction. Then either $ c $ or $ \mu $ is a symmetrical operator, or dyadic, as Willard Gibbs calls it.' (\cite{EM2}, 518). When comparing electromagnetic and elastic phenomena, Heaviside introduced also strain as pure rotation and, explicitly following Gibbs, he introduced the dyadical form of linear operators (\cite{EM2}, 526). Finally, by analysing the `Transformation of Characteristic Equation by Strain' for the electric and the magnetic field, the \textit{curl} of the operator $ c^{-1} $ appeared (\cite{EM2}, 531), namely [\textbf{H}]:
\begin{equation}\label{curl-vec-H}
-\mu \ddot{\vec{H}} = V\nabla c^{-1} V\nabla\vec{H}\; ,
\end{equation}
where Heaviside's $ \vec{H} $ `magnetic force' is the magnetic field. Once again, the right hand side of equation [\ref{curl-vec-H}] can be interpreted as the \textit{curl} of the operator $ c^{-1} $, i.e. [\textbf{H}] $  V\nabla c^{-1} $, applied to the vector [\textbf{H}] $ V\nabla\vec{H} $ or as the vector product between $ \nabla $ and the vector $ c^{-1}V\nabla\vec{H} $. As we already recalled, unlike Gibbs, Heaviside never discussed the \textit{curl} of a dyadic as an abstract mathematical object. Furthermore, as emerges from equations [\ref{eq-curl-E}] and [\ref{curl-vec-H}], Heaviside always applied it to a vector. Even if he never discussed the \textit{curl} of a dyadic, Heaviside recognised the importance of the idea in this context, because soon after, by criticising the `Quaternionic Innovations' in a paper published by \textit{Nature} at the beginning of 1894, the author considered Gibbs's advances in the theory of dyadics, the stress operator $ \phi $ and the \textit{nabla} operator and  emphasised: `See Gibbs's ``Elements of Vector Analysis'' (1881-4) for the direct product of $ \nabla $ and $ \phi $. (Also for the skew product, a more advanced idea; \textit{it, too, is a physically useful result}.)' [emphasis added] (\cite{EM3}, 512). This is another statement proving the impact that Gibbs's work had on Heaviside in the context of linear operators, but, in addition, from our point of view, in this statement Heaviside was implicitly referring to the physical meaning of $ curl\phi $ which we discussed in this section.

In the development of the theory of elasticity, the \textit{curl} of a tensor emerged from the so-called \textit{compatibility conditions} \cite{Dahan}. They are sufficient and necessary conditions to be imposed on a generic strain tensor, in order to correspond to a unique and continuous displacement field (\cite{Irgens}, 148). These conditions were published for the first time in 1861 by Adh\'emar Jean Claude B. de Saint-Venant (\cite{hist-of-elast}, 74) and they were developed in the context of elastic ether theory by Eugenio Beltrami in 1886 (\cite{Tazzioli}, 23). From a modern point of view, in the compatibility conditions the \textit{curl} is applied two times to the strain tensor $ \varepsilon $. By denoting with $ \varepsilon_{kl} $ its components, using index notations the compatibility conditions read (\cite{Irgens}):
\begin{equation}\label{Compat-conditions}
curl(curl\,\varepsilon )=0 \qquad\text{i.e.}\qquad\epsilon_{ikn}\partial_i\left( \epsilon_{jlm}\partial_j\varepsilon_{kl}\right)  = 0\; .
\end{equation}
Due to the antisymmetric properties of the alternating tensor, equations (\ref{Compat-conditions}) are equivalent to the following conditions (\cite{Irgens}, 148):
\begin{equation}\label{Compat-conditions-2}
\partial_{i}\partial_{j}\varepsilon_{kl}-\partial_{i}\partial_{l}\varepsilon_{kj} +\partial_{k}\partial_{j}\varepsilon_{ij}-\partial_{k}\partial_{j}\varepsilon_{il} = 0\; .
\end{equation}
Both Saint-Venant and Beltrami used this equivalent form, eq. [\ref{Compat-conditions-2}], but they did not use the index notation and the role of the \textit{curl} did not emerge explicitly. The importance of the index notation was emphasised only in the 1900s in order to popularize the absolute differential calculus by emphasising the importance of its application, e.g. in the context of the theory of elasticity \cite{Ricci-Levi-Civita}. Indeed, Gregorio Ricci and Tullio Levi-Civita opened their paper by quoting the following statement of Henri Poincar\'e: ``a good notation has the same philosophical importance as a good classification in natural sciences''\footnote{See \cite{Hermann} for a commented English translation.} (\cite{Ricci-Levi-Civita}, 125). But also in this paper, the \textit{curl} of a tensor did not emerge yet. The translation of the theory of elasticity in the language of absolute differential calculus continued with Synge \cite{Synge}, but the \textit{curl} of the strain tensor would emerge later, when the compatibility conditions would be rewritten in curvilinear coordinates (\cite{Gurtin}, 40).

Also in the context of vector calculus, the occurrence of the curl of a tensor in the compatibility conditions would have been highly recognizable, even if all the necessary ingredients had been present in Gibbs's booklet. Indeed, the equivalence between eq. [\ref{Compat-conditions}] and [\ref{Compat-conditions-2}] follows from the correspondence, in three dimensions, between the \textit{curl} of a vector and the anti-symmetrised gradient of the vector, which is a rank-2 tensor. Gibbs did not discuss this correspondence explicitly in his booklet, but it can be inferred by the following considerations. In the main text, Gibbs noticed that `every dyadic [$ \Phi $] may be divided in two parts' (\cite{Gibbs-pamphlet}, 51), i.e. the symmetric and the antisymmetric one. Its antisymmetric part $ \displaystyle \frac{1}{2}\left\lbrace \Phi - \Phi_C \right\rbrace  $ in Gibbs's notation, can be rewritten as follows:
\begin{equation}\label{correspondence}
\frac{1}{2}\left\lbrace \Phi - \Phi_C \right\rbrace = \mathcal{I}\times\left[ -\frac{1}{2}\Phi_{\times}\right] \; ,
\end{equation}
where $ \mathcal{I} $ and $ \Phi_{\times} $ have been defined by equation [\ref{id-dyadic}] and [\ref{phi-cross}] respectively and where, as already recalled, the conjugate dyadic $ \Phi_C $ is represented by the transposed matrix. If the dyadic is the gradient of a vector, i.e.  $\Phi_{ij} = \partial_i \alpha_j $ using index notation, hence, the vector $ \Phi_{\times} $ is the $ curl $ of the vector $ \alpha $. Equation [\ref{correspondence}] establishes the correspondence between the \textit{curl} of vector $ \alpha $ and the antisymmetric dyadic $ \partial_i \alpha_j - \partial_j \alpha_i $, because $ \Phi_{\times} = 0 $ is equivalent to $ \partial_i \alpha_j - \partial_j \alpha_i =0$. As far as we know, Heaviside never noticed it explicitly. A similar correspondence can be established between the dyadic $ curl\,\phi $ and the antisymmetric part of the gradient of $\phi $, which is a triadic. These correspondences are commonly used nowadays also, and the \textit{curl} of a second rank tensor $ \phi $ can be identified with $ \partial_{i}\phi_{jk}-\partial_{k}\phi_{ji} $. Therefore, after receiving Gibbs's booklet, Heaviside was aware of this correspondence and he would have inferred the similar correspondence for tensors. But we did not find any evidence of this fact in Heaviside's published work.

\section{Divergence and Stokes theorems generalised}\label{Green-Stokes}
\begin{figure}[ht!]
	\centering
	\includegraphics[width=90mm]{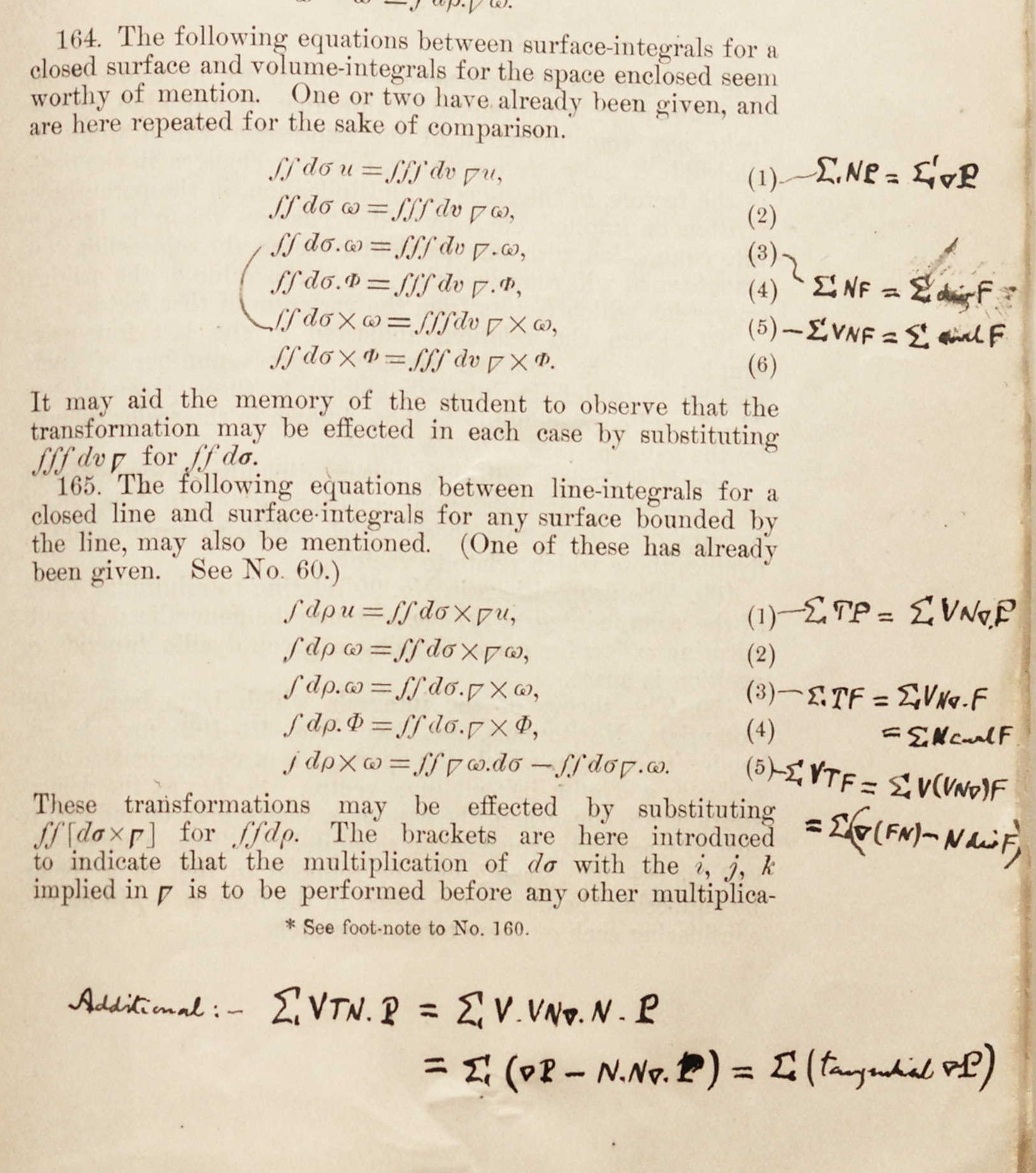}
	\caption{\textbf{In the left margin:} Heaviside's translation of Gibbs's formulas for surface- to volume- integrals, upper three lines, and for line- to surface-integrals, lower three lines. \textbf{At the bottom of the page:} Heaviside's new formula for line- to surface-integrals (p. 67)}
	\label{divergence-stokes}
\end{figure}

In this section, we shall discuss the last annotations taken  by Heaviside, which consist of two sets of translations of the numbered equations on page 67, see Fig. \ref{divergence-stokes}, placed in the right margin and at the bottom of the page.

First, we shall analyse the first three lines, relating surface-integrals to volume-integrals, then we shall consider the other annotations relating line-integrals to surface-integrals. Heaviside translated only some of the formulas that Gibbs presented. Unlike Gibbs, Heaviside used neither the integral symbol nor the differential symbol. Instead of the integral symbol, he used a summation symbol, while the differentials were implied. This choice made it difficult to understand, at the beginning of our research, whether the integrals at the bottom of page 67 were line-, surface- or volume-integrals, see Fig. \ref{divergence-stokes}. By comparing the annotation with Heaviside's published work, we found the formula at the bottom of the page in \textit{Electromagnetic Theory}. Hence, we inferred that it relates suface- to line-integrals. As a consequence, as already said, we marked August 1892 as the end of the period when the annotations were made.
 
\subsection{From surface to volume integrals}
The first set of annotations, placed in the right margin of page 67, is concerned with the Divergence theorem and its generalisation for rank two and three tensors.
\begin{figure}[ht!]
	\centering
	\includegraphics[width=90mm]{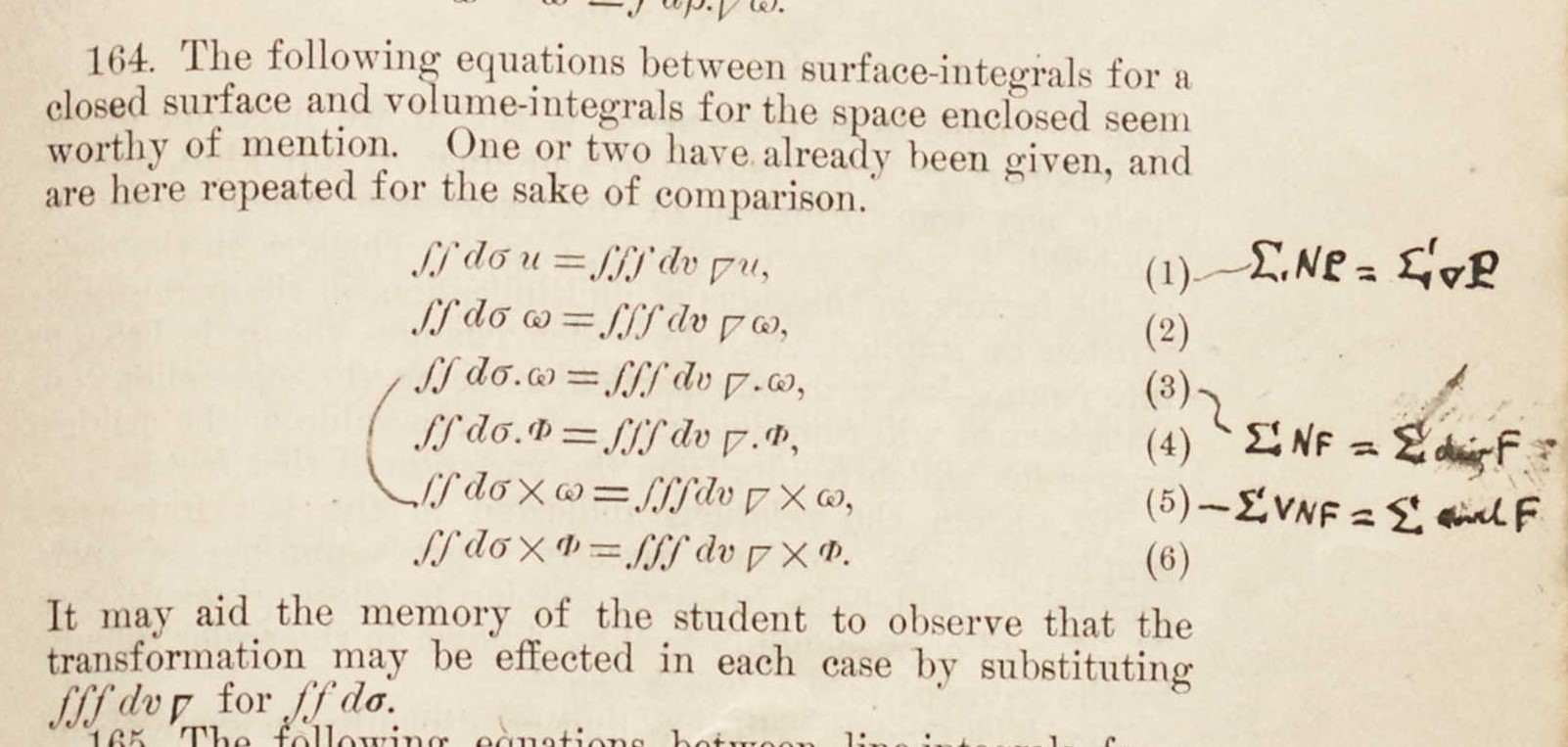}
	\caption{\textbf{In the left margin:} Heaviside's translation of Gibbs's formulas for surface- to volume-integrals}
	\label{divergence}
\end{figure}
The Divergence theorem was discovered before Gibbs and Heaviside and the beginning of this story was investigated by Charles Stolze in \cite{Stolze}. Stolze observed that in 1893 `Heaviside published his famous work on electromagnetic theory' (\cite{Stolze}, 441), i.e. \cite{EM1}, and then pointed out: `It also seems to be the first explicit reference to the divergence
theorem as such. The term divergence had been used earlier, but apparently never in direct reference to the theorem.'  (\cite{Stolze}, 441). We disagree with Stolze, because on December 2, in 1882, a paper by Heaviside was published in the \textit{Electrician}, where the author explicitly considered the `Theorem of Divergence' (\cite{ELP1}, 209). Heaviside presented proof of the theorem emphasising that he had read about it `in a German work' (\cite{ELP1}, 208), without quoting which. A formulation of the theorem in vector notation was published in \textit{Electromagnetic Theory} (\cite{EM1}, 190, eq. (145)) and it is similar to the annotation made on Gibbs's booklet\footnote{As already noticed, Gibbs's notation looks like modern formulation.}, see eq. [\ref{div3}].

In Gibbs's notation, the vector character of the differentials is implied by the use of Greek letters. In the following we shall make explicit both the vector character by using Clarendon's notation and the tensor product\footnote{For example, Gibbs's vector $ \omega $ will be denoted by $\boldsymbol{\omega} $ and the indeterminate product of the l.h.s of eq. (2) in Fig. \ref{divergence}, i.e. $d\sigma\,\omega  $, reads $ d\vec{\sigma}\otimes\omega $.}. We introduced the unit normal $ \vec{N} $ to the surface considered, like in Heaviside's annotations, therefore Gibbs's surface element $ d\boldsymbol{\sigma} $ reads $ \vec{N}\; dS $. Hence, Fig. \ref{divergence} reads:
\begin{eqnarray}
	& \text{Gibbs} & \qquad\qquad\qquad\qquad\qquad\qquad\text{Heaviside}\nonumber \\
	\nonumber\\
	\int \vec{N}u\; dS & = & \int \nabla u\; dv \qquad\qquad\qquad\quad \sum\;\vec{N}P = \sum\;\nabla P     \label{div1}  \\
	\int \vec{N}\otimes\boldsymbol{\omega} \; dS & = & \int \nabla \otimes\boldsymbol{\omega}\; dv \qquad\qquad\qquad \label{div2} \\
	\int \vec{N}\boldsymbol{\cdot}\boldsymbol{\omega} \; dS & = & \int \nabla \boldsymbol{\cdot}\boldsymbol{\omega}\; dv \qquad\qquad\qquad \sum\;\vec{N}\vec{F} = \sum\;\nabla\vec{F}      \label{div3}\\
	\int \vec{N}\boldsymbol{\cdot}\Phi\; dS & = & \int \nabla\boldsymbol{\cdot}\Phi \; dv \qquad\qquad\qquad \label{div4}\\
	\int \vec{N}\times\boldsymbol{\omega} \; dS & = & \int \nabla \times\boldsymbol{\omega}\; dv \qquad\qquad\quad \sum\; V\vec{N}\vec{F} = \sum\; curl\,\vec{F}   \label{div5}   \\
	\int \vec{N}\times\Phi\; dS & = & \int \nabla\times\Phi \; dv \qquad\qquad\qquad \label{div6}
\end{eqnarray}
In equation [\ref{div1}], Heaviside changed the name of the scalar function $ u $ into $ P $, like the pressures introduced in continuum mechanics, while in equations (\ref{div3}) and (\ref{div5}) Heaviside changed the name of Gibbs's vector $ \boldsymbol{\omega} $ into $ \vec{F} $. Equation (\ref{div3}) is the Divergence theorem. 

Why did Heaviside not translate equations (\ref{div2}), (\ref{div4}) and (\ref{div6})? Did Heaviside already know the expressions he translated? Like Ido Yavetz pointed out (\cite{Heav-biblio-Yavez}), in the first volume of his \textit{Electromagnetic Theory}, Heaviside presented a generalisation of the Divergence theorem. Gibbs's equations (\ref{div4}) and (\ref{div6}) can also be regarded as generalisations of the Divergence theorem. Are there any differences between Heaviside and Gibbs's statements? Below, we shall address these questions. 

The so-called Divergence theorem relates the surface integral of a continuously differentiable vector field to the integral over a volume of its divergence. Like Gibbs, Heaviside considered also the gradient theorem, i.e. eq. [\ref{div1}], where a scalar function, instead of a vector field, is involved. But the main novelty of Heaviside's approach, as pointed out also by Yavetz, lies in the fact that, in \textit{Electromagnetic Theory}, he considered a generalisation where scalar and vector arbitrary integrand functions of the unit vector field of the surface itself are involved. In order to formulate his generalisation, Heaviside introduced an additional hypothesis regarding the parity of the integrand function. Even if this generalisation does not emerge from the annotations, we will briefly describe Heaviside's argument, like Yavetz, and then we shall add some new comments.
Heaviside's statement can be summarised as follows:\\
\textbf{Theorem:} \textit{Let $ \mathcal{S} $ be a closed surface and $ V $ its enclosed volume. Let $ \vec{N} $ be the outward unit normal to the surface. Given an arbitrary scalar (vector) function of the coordinates and of the unit normal} $ \mathcal{S} $, \textit{namely} $ F\left( x,y,z; \vec{N}\right)  $ ($ \vec{F}\left( x,y,z; \vec{N}\right)  $)\textit{, the integral over the surface itself can be rewritten as a volume integral of another function,} $ f\left( x,y,z; \vec{N}\right)  $ ($ \vec{f}\left( x,y,z; \vec{N}\right)  $)\textit{, provided that} $ F\left( x,y,z; \vec{N}\right)  $ ($ \vec{F}\left( x,y,z; \vec{N}\right)  $)\textit{ is an odd function of} $ \vec{N} $ (\cite{ELP1}, 190). 

Using modern language, the two statements, for scalar and vector functions respectively, read:
\begin{eqnarray}
\int_\mathcal{S} F\left( x,y,z; \vec{N}\right) dS  &=& \int_{V}  f\left( x,y,z; \vec{N}\right) dv\; , \label{th-div-H1}\\
\int_\mathcal{S} \vec{F}\left(  x,y,z; \vec{N} \right) dS &=& \int_{V}  \vec{f}\left( x,y,z; \vec{N}\right) dv\; .\label{th-div-H2} 
\end{eqnarray}
where $ V $ is the volume enclosed by $ \mathcal{S} $. It is worth noting that Heaviside did not give a rigorous proof of the statement, by modern standard, but he sketched it as follows. First, Heaviside considered the statement of the Divergence theorem. `Consider the summation $ \sum\vec{N}\vec{D} $ of the normal component of a vector $ \vec{D} $ over any closed surface.' (\cite{EM1}, 190). As already said, the differential $ dS $ is implied in Heaviside's notation and, in addition, it must be noticed that Heaviside's `vector $ \vec{D} $' should be regarded as a \textit{vector field}. He emphasised that in considering the usual Divergence theorem `the vector $ \vec{D} $, to which it is applied, admits of finite differentiation.' (\cite{EM1}, 190). Then,  the proof continues as follows. `Divide the region enclosed into two regions. Their bonding surfaces have a portion in common. If, then, we sum up the quantity $ \vec{N}\vec{D} $ for both regions (\textit{over their boundaries, of course}), the result will be the original $ \sum\vec{N}\vec{D} $ for the complete region.' [emphasis added] (\cite{EM1}, 190). Heaviside specified that the integration must be intended over the surface, where the unit normal is well defined. `The normal is always to be reckoned positive outwards from a region so that on the surface common to the two smaller regions, $ \vec{N} $ is $ + $ for one and $ - $ for the other region [...]' (\cite{EM1}, 190). Hence, the contribution of the common surface to the total integral is zero. `Since the process of division may be carried on indefinitely [...] the summation $ \sum\vec{N}\vec{D} $ for the boundary of any region equals the sum of the similar summations applied to the surfaces of the similar summations of all the elementary regions into which we may divide the original. That is,
\begin{equation}
\sum\vec{N}\vec{D} = \sum \phi\left( \vec{D}\right) \; ,\nonumber
\end{equation}
where, on the right side, we have a volume-summation whose elementary $ \phi\left( \vec{D}\right) $ part is the same quantity $ \sum\vec{N}\vec{D} $ as before, belonging now, however, to the elementary volume in question. We have already identified $ \phi\left( \vec{D}\right) $ with the divergence of $ \vec{D} $.' (\cite{EM1}, 191). Indeed, on page 189, Heaviside had considered the unit cube case, and he had obtained the divergence of the vector field. Having stated this premise, the author introduced his generalisation. Heaviside emphasised that `the validity of the process whereby we pass from a surface- to a volume-summation, depends solely upon the quantity summed up, viz., $ \vec{N}\vec{D} $, changing its sign with $ \vec{N} $.' (\cite{EM1}, 191). Then, he concluded: `We may therefore at once give the Divergence theorem a wide extension, making it [...] take this form\footnote{The number of the original equation has been changed.}:
\begin{equation}\label{div-th-Heaviside}
\sum F\left( \vec{N} \right) = \sum f(\vec{N})\; 
\end{equation}
Here, on the left side, we have a surface-, and on the right side a volume-summation. \textit{The function} $ F(\vec{N}) $, where $ \vec{N} $ is the outward normal,\textit{ is any function which changes sign with }$ \vec{N} $. The other function $ f(\vec{N} ) $, the element of the volume-summation, is the value of $ \displaystyle \sum F\left( \vec{N} \right) $ for the surface of the element of volume.'[emphasis added] (\cite{EM1}, 191). Equation [\ref{div-th-Heaviside}] summarises eq. [\ref{th-div-H1}] and [\ref{th-div-H2}]. Finally, Heaviside also gave the Cartesian form for $ f\left( x,y,z; \vec{N}\right)  $ and $ \vec{f}\left( x,y,z; \vec{N}\right)  $ when the surface is `a cubical element' (\cite{EM1}, 191), namely [\textbf{H}]:
\begin{equation}\label{f-Heaviside}
 f\left( x,y,z; \vec{N}\right) = \nabla_1 F\left( \vec{i}\right) +\nabla_2 F\left( \vec{j}\right)  +\nabla_3 F\left( \vec{k}\right)  \; ,
\end{equation}
\begin{equation}\label{f-vec-Heaviside}
\vec{f}\left( x,y,z; \vec{N}\right) = \nabla_1 \vec{F}\left( \vec{i}\right) +\nabla_2 \vec{F}\left( \vec{j}\right)  +\nabla_3 \vec{F}\left( \vec{k}\right)  \; ,
\end{equation}

From Heaviside's point of view, equations (\ref{div1}), (\ref{div3}) and (\ref{div5}) can be regarded as applications of the general statement. Indeed, they appeared as examples in the \textit{Electrician} between the end of March and the beginning of April 1893: 
Heaviside described some examples of his general statement, which correspond precisely to the formulas he translated in Gibbs's booklet. Examples $ a) $ and $ b) $, see (\cite{EM1}, 193), correspond to eq. [\ref{div1}]; examples $ c) $ and $ d) $ (\cite{EM1}, 193 and p. 194) to eq. [\ref{div3}] and the last example $ e) $  (\cite{EM1}, 194) corresponds to eq. [\ref{div5}]. In this part of \textit{Electromagnetic Theory}, Heaviside did not mention explicitly Gibbs's work, but the generalisation of the Divergence theorem was published on March 25, 1892, and, as already mentioned in section \ref{reciprocal}, he had already commented on Gibbs's booklet on November 13, 1891. In addition, in all of Gibbs's formulas translated by Heaviside the integrand, i.e. $ F\left( \vec{N} \right) $, is an odd function of the normal unit vector $ \vec{N} $. Hence, from our point of view, it is highly probable that Heaviside started to elaborate his generalisation of the theorem after reading Gibbs's booklet.

Heaviside did not discuss the properties of the unit vector field $ \vec{N}\left( \vec{x}\right)  $. If the surface is a smooth orientable manifold, the Gauss map provides a continuous differentiable normalized vector field over the whole volume enclosed by the surface\footnote{Except for a set of null-measure, e.g. a spere of radius $ R $.}. Hence, Gibbs's generalisation of the Divergence theorem for tensors holds, see equation [\ref{div4}], and the additional hypothesis on the parity of the integrand is not needed. Therefore, for smooth manifolds, Heaviside's theorem can be regarded as a particular case of Gibbs's theorem. Indeed, Heaviside's theorems eq. [\ref{th-div-H1}] and [\ref{th-div-H2}] can be obtained as follows. By setting $ \boldsymbol{\omega}=F\left( \vec{N}\right) \vec{N}$ in eq. [\ref{div3}], it is equivalent to eq. [\ref{th-div-H1}], with $ f\left( x,y,z; \vec{N}\right) = \nabla\boldsymbol{\cdot}\left( F\vec{N}\right) $. Analogously, by choosing $ \Phi = \vec{N}\otimes \vec{F}\left( \vec{N}\right) $, eq. [\ref{div4}] is equivalent to eq. (\ref{th-div-H2}), with $ 
\vec{f}\left( x,y,z; \vec{N}\right) =\nabla\boldsymbol{\cdot}\left( \vec{N}\otimes\vec{F} \right) $.

What is the role of the additional hypothesis? In order to address this question, we make the following remarks. First, in \textit{Electromagnetic Theory} the vector field $ \vec{N} $ lives on the surface. This means that, from the three-dimensional point of view, it does not correspond to a continuously differentiable vector field. Indeed, as we discuss in appendix, section \ref{parity}, the unit normal vector field should be constructed with the help of Dirac's delta function $ \delta (x) $, which is a ``generalised function'', i.e. an operator. It is worth noting that Heaviside would introduce explicitly the delta function as the derivative of the step function three years later, in 1895, in the context of operational theory (\cite{EM2}, 55), and, as Jesper L\"utzen pointed out, he would also introduce its fundamental property, namely $ \displaystyle \int_{-\infty}^{+\infty}\delta(x-a)f(x)dx=f(a) $ (\cite{Lutzen}). But as far as we are aware, he never used it in the context of Divergence theorem. Second, in his booklet, Gibbs discussed the Divergence theorem for vector fields that have singular points in the origin, e.g. the electric field generated by a point charge (\cite{Gibbs-pamphlet}, 35), but no annotations are present in this part of the pamphlet. Yavetz expanded Heaviside's explanation in order to show how to obtain eq. [\ref{f-Heaviside}] or (\ref{f-vec-Heaviside}) and he mentioned Heaviside's additional hypothesis. In appendix, section \ref{parity}, we investigate in more detail its role and we show that if we allow the function $ F(\vec{N}) $, or $ \vec{F}(\vec{N}) $, to be differentiable in the distribution sense, the additional hypothesis is needed in order to obtain Heaviside's specific form for $ f $ and $ \vec{f} $, i.e. eq. [\ref{f-Heaviside}] and (\ref{f-vec-Heaviside}). Therefore, the additional hypothesis permitted Heaviside to avoid the use of the delta function for the simplest case of a cubical element and to obtain a formula involving the divergence operator applied to $ F(\vec{N}) $, or $ \vec{F}(\vec{N}) $, which is well defined in the case of cubical element. Third, nowadays, the validity of equations (\ref{div1}) and (\ref{div5}) can be proved as follows. Let $ \vec{A} $ be a constant vector: after making the scalar product between $ \vec{A} $ and the left-hand side of eq. [\ref{div1}] or eq. [\ref{div5}], the Divergence theorem and the properties of the mixed product are sufficient to end with the proof. This kind of proofs is based on the assumption that the operator $ \nabla $ can be formally manipulated like a vector into vector formulas. This assumption could provide shortcuts for manipulating expressions, but identifying the \textit{nabla} operator with a vector could produce incorrect manipulations, as indeed it did, as Chen-To Tai pointed out in \cite{Tai-1}, \cite{Tai-2} and \cite{Tai-3}. In particular, in \cite{Tai-1}, the author emphasised how Gibbs never considered the \textit{nabla} operator as a formal vector and that this tradition started with Wilson's book (\cite{Tai-2}: p. 8). Let us now come back to equations (\ref{div1}) to (\ref{div6}). Gibbs emphasised that `it may aid the memory of students to observe that the transformation may be effected by substituting $ \displaystyle \int dv\,\nabla $ for $\displaystyle \int dS\vec{N} $.'\footnote{We changed Gibbs's notation like in equations (\ref{div1})-(\ref{div6}).} (\cite{Gibbs-pamphlet}, 62). Also Heaviside noticed the same shortcut, but he had a different attitude to it. Indeed, in \textit{Electromagnetic Theory}, he remarked: `Observe that in all of the above examples,
\begin{equation}
\sum \, F\left( \vec{N}\right) = \sum \, F\left( \nabla\right)\; ,\nonumber
\end{equation} 
when we pass from a closed surface to the enclosed region [...]. But I cannot recommend anyone to be satisfied with such symbolism alone. It is much more instructive to go more into details [...] and see how the transformations occur, bearing in mind the elementary reasoning upon which the passage from one kind of summation to another is based[...]'\footnote{The number of the equation has been omitted} (\cite{EM1}, 199). In spite of this fact, as we shall see in the following section, Heaviside used it extensively. 

Before proceeding, we point out that, as far as we know, Heaviside never used the integral calculus for tensors in his published work. Hence, from our point of view, Heaviside did not translate equations (\ref{div2}), (\ref{div4}) and (\ref{div6}), because he did not see whether they could have a direct impact on his work.

\subsection{From line to surface integrals}
The second set of annotations in the right margin of page 67 concerns with Stokes's theorem and its generalisation. Heaviside called it `the Theorem of Version' at the end of 1882 (\cite{ELP1}, 211), but he would start to call it Stokes's theorem ten years later (\cite{EM1}, 193).
\begin{figure}[ht!]
	\centering
	\includegraphics[width=90mm]{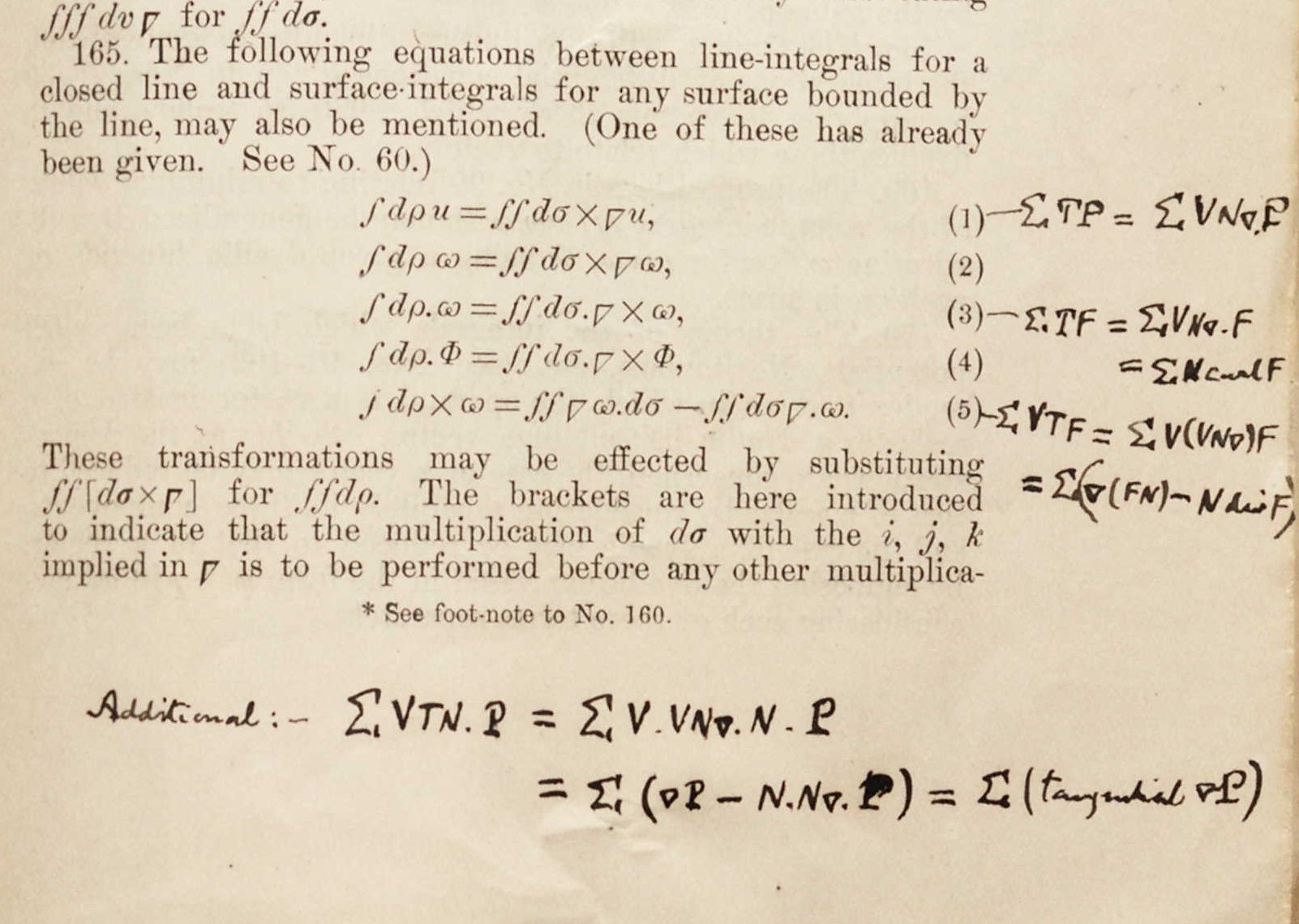}
	\caption{\textbf{In the left margin:} Heaviside's translation of Gibbs's formulas for line- to surface-integrals. \textbf{At the bottom of the page:} Heaviside's new formula for line- to surface-integrals}
	\label{stokes}
\end{figure} 

Like in the preceding section, below we rewrite both Gibbs's text, by making clear the vector character and the tensor product, as well as Heaviside's annotations. The formulas relate integrals along closed curves to those over the enclosed surface. We introduced the tangent vector to the curve $ \vec{T} $ and Gibbs's line element $ d\boldsymbol{\rho} $ has been substituted with $ \vec{T}dr $.
\begin{eqnarray}
& \text{Gibbs} & \qquad\qquad\qquad\qquad\qquad\qquad\qquad\text{Heaviside}\nonumber \\
\int \vec{T}u\; dr & = & \int \vec{N}\times\nabla u\; dS \qquad\qquad\qquad \sum\;\vec{T}P = \sum\;V\vec{N}\nabla . P     \label{stokes1}  \\
\int \vec{T}\otimes\boldsymbol{\omega} \; dr & = & \int \vec{N}\times\nabla \otimes\boldsymbol{\omega}\; dS  \label{stokes2} \\
\int \vec{T}\boldsymbol{\cdot}\boldsymbol{\omega} \; dr & = & \int \vec{N}\boldsymbol{\cdot}\nabla \times\boldsymbol{\omega}\; dS \qquad\qquad\quad \sum\;\vec{T}\vec{F} = \sum\;V\vec{N}\nabla .\vec{F} \nonumber\\
&  & \qquad\qquad\qquad\qquad\qquad\qquad\qquad\qquad = \sum\vec{N}\, curl\, \vec{F}     \label{stokes3}\\
\int \vec{T}\boldsymbol{\cdot}\Phi\; dr & = & \int \vec{N}\boldsymbol{\cdot}\nabla\times\Phi \; dS \qquad\qquad \label{stokes4}\\
\int \vec{T}\times\boldsymbol{\omega} \; dS & = & \int \nabla\boldsymbol{\omega}\boldsymbol{\cdot}\vec{N} \; dS - \int \vec{N}\nabla\boldsymbol{\cdot}\boldsymbol{\omega} dS \quad\quad \sum\; V\vec{T}\vec{F} = \sum\; V\left( V\vec{N}\nabla\right)\,\vec{F}  \nonumber\\
&  & \qquad\qquad\qquad\qquad\qquad\,\quad\quad\qquad = \sum\left( \nabla\left( \vec{F}\vec{N}\right) - \vec{N}\, div\, \vec{F}\right)   \label{stokes5}   
\end{eqnarray}

Again, Gibbs emphasised that these `transformations may be effected by substituting $ \displaystyle\, \int dS\; \vec{N}\times\nabla\quad $  for  $\quad \displaystyle \int dr\; \vec{T}\; $.'\footnote{Our usual notation has been applied. In the original text a misprint is present, because the integral symbol has been wrongly doubled for the line integral $ \displaystyle \int \vec{T}dr $.} (\cite{Gibbs-pamphlet}, 62). As already said, Heaviside also noticed the same shortcut, and in these annotations he used it. Indeed, by considering  equations (\ref{stokes3}) and (\ref{stokes5}), he firstly substituted $ \vec{T} $ for [\textbf{H}] $ V\vec{N}\nabla $ and then wrote an equivalent form for the formulas. Heaviside used the same trick at the bottom of the page where the following annotation appears:
\begin{eqnarray}
\text{Additional:}\quad \sum V\vec{T}\vec{N}. P & = & \sum V.V\vec{N}\nabla . \vec{N} . P \label{stokes6a}\\
& = & \sum \left( \nabla P - \vec{N} . \vec{N}\nabla . P\right) = \sum \left( \text{tangential} \nabla P\right) \label{stokes6b}\; ,
\end{eqnarray}
where $ P $ is a scalar function. Equations (\ref{stokes6a}) and (\ref{stokes6b}) relate the integral along a circuit, i.e. the first sum in eq. [\ref{stokes6a}], with integrals over the surface enclosed by the circuit, i.e. the other sums. Heaviside used Gibbs's shortcut suggestion in order to obtain the second sum of eq. [\ref{stokes6a}] and then wrote two equivalent formulas in the second line, eq. [\ref{stokes6b}]. In the published version of this annotation, Heaviside changed the order of the integrals: `In general, therefore,
\begin{equation}
\sum V\vec{T}\vec{N}. P  =
\sum \left( \nabla P - \vec{N} . \vec{N}\nabla . P\right)= \sum \nabla_s P = \sum V.V\vec{N}\nabla . \vec{N} . P \; .
\end{equation}
[...] The last form [...] involves the \textit{transformation formula} [...]'[emphasis added] (\cite{EM1}, 198). Heaviside's \textit{transformation formula} is the double cross-product law.

In \textit{Electromagnetic Theory}, by discussing examples of line- to surface-integrals, Heaviside presented the same formulas he translated, i.e. equations (\ref{stokes1}) -- (\ref{stokes5}). But coherently with his attitude toward the shortcut method, the reasoning followed by the author in presenting the derivations is reversed. A single example will be sufficient. Let us consider eq. [\ref{stokes3}], i.e. Stokes's theorem. First, Heaviside gave the proof of the following identity, namely [\textbf{H}] $ \displaystyle  \sum\;\vec{T}\vec{F} = \sum\vec{N}\, curl\, \vec{F}  $, by summing as usual the contribution of elementary circuits. Then, he noticed that it can be rewritten in the following equivalent form [\textbf{H}] $ \sum\;V\vec{N}\nabla .\vec{F} $. Heaviside had never published any of these formulas, therefore the annotations on Gibbs's pamphlet seem to suggest that he learned them by studying Gibbs's booklet. 

In analogy with the path followed for the Divergence theorem, Heaviside proposed a generalisation of the Stokes's theorem. From Heaviside's point of view, again, equations ($ \ref{stokes1} $), ($ \ref{stokes3} $) and ($ \ref{stokes5} $) can be obtained from a general statement, which should now involve an arbitrary odd function of the tangent vector $ \vec{T} $. All the remarks made in the preceding section could be repeated for the generalisation of Stokes's theorem, as we briefly show in appendix \ref{parity}.

In the same section of \textit{Electromagnetic Theory}, Heaviside also presented his additional equations (\ref{stokes6a}) and (\ref{stokes6b}) as further example of the generalised statement (\cite{EM1}, 198). Is there any correspondence between Heaviside's addendum, i.e. equations (\ref{stokes6a}) and (\ref{stokes6b}), and Gibbs's formulas? Did Heaviside invent it or did he find it in Gibbs's booklet? Below, we shall address these last two questions. First, we rewrite  eq. [\ref{stokes6a}] by using Gibbs's notation, namely:
\begin{equation}\label{stokes6-bis}
\int \vec{T}\times\vec{N} P \; dr = \int \left( \vec{N}\times\nabla\right) \times\vec{N} P  \; dS\; .
\end{equation}
Then, we point out that equation [\ref{stokes6-bis}] could be regarded as a particular case of eq. [\ref{stokes4}]. Indeed, it can be obtained by using Gibbs's dyadic $ \Phi = \left\lbrace \mathcal{I} \times\vec{N}P\right\rbrace  $, which acts on the left on the vector $ \vec{T} $ as follows: $ \vec{T}\boldsymbol{\cdot}\left\lbrace \mathcal{I} \times\vec{N}P\right\rbrace \overset{def}{=}\vec{T}\times\vec{N}P $ (\cite{Gibbs-pamphlet}, 48). By using index notation, the same property reads $\Phi_{ij}= \left\lbrace \mathcal{I} \times\vec{N}P\right\rbrace _{ij} =\delta_{il}\epsilon_{lkj}N_k P= \epsilon_{ikj}N_k P  $. It is worth noting that Heaviside made no annotations on the pages where this example of Gibbs's dyadic is presented. The equivalence between Heaviside's eq. [\ref{stokes6-bis}] and Gibbs's eq. [\ref{stokes4}] follows easily by using index notation. Equation [\ref{stokes4}] reads 
\begin{equation}
\int T_i\Phi_{ij}\;  dr     =  \int N_i\epsilon_{ilm}\partial_{l}\Phi_{mj} \; dS \; ,
\end{equation} 
and by using the components form for $ \Phi_{ij} $, it can be rewritten as follows:
\begin{eqnarray}
\int T_i\left[  N_k P\epsilon_{ikj}\right] \; dr &=& \int N_i\epsilon_{ilm}\partial_{l}\left[ \epsilon_{mkj}N_k P\right]  \; dS \label{components1}\\
&=& \int \left[ N_i\epsilon_{ilm}\partial_{l}\right] \epsilon_{mkj}\left( N_k P \right)  \; dS \label{components2}
\end{eqnarray} 
where we simply moved the position of the squared parentheses in order to make evident that in eq. [\ref{components2}] the components of the r.h.s of eq. [\ref{stokes6-bis}] appear. Finally, we consider eq. [\ref{stokes6b}]. By using the Levi-Civita property $\displaystyle \epsilon_{ilm}\epsilon_{mkj} = \delta_{ik}\delta_{lj} - \delta_{lk}\delta_{ij} $, it is easy to see how eq. [\ref{components2}] reduces to eq. [\ref{stokes6b}]:
\begin{eqnarray}
\int  N_i \left( \delta_{ik}\delta_{lj} -  \delta_{lk}\delta_{ij}\right) \partial_{l}\left( N_k P \right)  \; dS &=& \int \left[  N_i \partial_{j}\left( N_i P \right)  - N_j \partial_{k}\left( N_k P \right) \right]  \; dS \nonumber\\
 &=& \int \left[  \partial_{j} P  - N_j N_k\partial_{k}  P \right]  \; dS \; ,\label{stokes-final}
\end{eqnarray}
where we used the constancy of $ \vec{N} $ in order to obtain the last line of eq. [\ref{stokes-final}]. Heaviside translated neither eq. [\ref{stokes2}], where the tensor product was involved, nor eq. [\ref{stokes4}], which is connected with his addendum formula. Therefore, from our point of view, it can be claimed that Heaviside obtained his addendum formula independently and that he did not consider it as part of Gibbs's results.

\section{Summary and Conclusions}\label{Conclusions}

In June 1888, Oliver Heaviside received by mail an unpublished pamphlet, which was written and printed between 1881 and 1884 by the American physicist Willard J. Gibbs. Before 1888, the two authors had developed independently the modern system of vectors, but with different notations: Heaviside adapted the British tradition of quaternionists, while Gibbs invented his own. The aim of our paper was to investigate if Gibbs's booklet had an impact on Heaviside's work. In section \ref{Gen-remarks}, we argued that the pamphlet received by Heaviside is now conserved at the \textit{Dibner Library for the History of Science and Technology}. Many annotations were made on the booklet. Hence, we inferred that Heaviside studied Gibbs's work very carefully. By comparing Heaviside's annotations and his published work, we argued that the annotations were made between June 1888 and April 1892.

In our main text, we presented the most relevant annotations. Heaviside expressed his admiration for Gibbs's contribution, but he also criticised, in his published work, some of the choices made by the American physicists. By analysing the annotations we selected, we argued that all the critiques expressed by Heaviside originated from his critical analysis of Gibbs's booklet. Heaviside criticised Gibbs's symbols for the scalar, vector and tensor product and decided to keep on using his own notation, because he derived it from the quaternionists tradition. Indeed he translated all of the formula he was interested in. Heaviside discussed the concept of the reciprocal of a vector, which he had already introduced following the quaternionist tradition, section (\ref{reciprocal}). Then, he criticised Gibbs's \textit{indeterminate product}, i.e. the tensor product, because he never considered it as an abstract product, section \ref{indeterminate}.

In the same section, we pointed out the impact that Gibbs had on Heaviside's work by discussing the annotations regarded with the topic of linear operators. Heaviside himself emphasised Gibbs's role in this area of Mathematics. From the annotations, it emerged that Heaviside particularly appreciated the concise method offered by the linear operator for solving vector equations formally. Also, it emerged that Heaviside learned to manage Gibbs's dyadics and that he introduced a new triadic, see section \ref{triadic-section}. As far as we know, Heaviside never discussed this abstract concept in his published works, but he used it implicitly by applying vectors to this operator, as we suggested in the same section. 

Heaviside introduced the concept of linear operators in the context of Maxwell's electromagnetism and by considering continuum mechanics. By studying Gibbs's booklet, he found some correspondences between the tools he used in the context of the linear operators and the mathematical objects defined in Gibbs's pamphlet, which gave a more abstract presentation of the subject. Heaviside was stimulated by this comparison and tried to understand the physical meaning of the \textit{curl} of a tensor in the context of continuum mechanics, a concept that he had never introduced before. Heaviside made two annotations, one inside the booklet and one on the back cover. We made a full transcription of the back cover annotation, because it was the most damaged part of the pamphlet, in section \ref{app1}. In section \ref{lin-operator-section}, we analysed his annotations step by step and discussed the emergence of this concept in Heaviside's published work and in the context of the theory of elasticity. 

The last section of our paper dealt with the generalisations of the Divergence and Stokes's theorem. Gibbs formulated a generalisation of the two statements for dyadics, while Heaviside considered arbitrary scalar and vector integrand functions depending on the normalised vector field normal to a closed surface, for the former theorem, and the unit tangent to a closed line for the latter. Heaviside formulated his generalisation after receiving Gibb's booklet. Hence, we investigated the connection between the two formulations. By comparing Heaviside's annotations with his published work, we argued that the formulation of his generalisations was stimulated by Gibbs's booklet. Furthermore, we showed that for continuously differentiable vector fields Gibbs's formulation is a generalisation of Heaviside's theorem. However, the British scientist, as far as we know, never realized. In addition, we considered the fact that in Heaviside's generalisation non-differentiable normal or tangent vector fields are involved and that Heaviside introduced an additional hypothesis on the parity of the integrand functions involved. In the main text, we suggested that these two facts are related, but we skipped technical details, which are discussed in appendix \ref{parity}. 

The following appendix, section \ref{app1}, contain a complete transcription of the most damaged part of Gibbs's booklet, where Heaviside discussed the meaning of the \textit{curl} of a tensor. In the last section, appendix \ref{others}, we present, for the sake of completeness, the annotations we did not discuss in the main text.

\section*{Acknowledgements}
The author is grateful to Giulio Peruzzi for his inspiring discussions; to Niccol\`o Guicciardini for having believed in this project and for his suggestions; to Kurt Lechner for his mathematical support and to Bruce Hunt for sharing his point of view on Heaviside's approach. The author would like to thank the anonymous referees for giving constructive comments which helped improving the organization and the quality of our paper. Finally, Lilla Vekerdy, the curator of the Dibner Library, and Morgan Aaronson deserve special acknowledgements for their invaluable support during the period spent in Washington DC. The financial support was provided by a three-months fellowship of the \textit{Dibner Library Resident Scholar Program}.All the photographs of our paper were provided by courtesy of the Dibner Library.
\begin{appendices}

\section{On Heaviside's generalisations of the Divergence theorem: some technical details.}\label{parity}
The Divergence theorem is usually formulated for continuously differentiable vector fields, its \textit{classical version}, however, as suggested in Gibbs's booklet, it holds for scalar and tensor fields also. Nowadays, we know that it can be generalised for  tempered distributions, by introducing the correct measure in order to carefully give sense to the integrals involved \cite{distribution1} \cite{distribution2}. By allowing the function of the theorem to be a function of the unit vector of the surface itself, the theorem can be applied in its classical version if the surface $ \mathcal{S} $ is an orientable differentiable manifold. Indeed, in this case the Gauss map  $ \vec{N}(\vec{x})  $ can be globally defined and it represents the unit normal vector field of the theorem, pointing outward as usual. Hence, Heaviside's request that the scalar, or vector, integrand be an odd function of $ \vec{N}(\vec{x})  $ is not needed. In the case of non-regular surfaces or in the case of a cubic surface, we would need the generalised version for distributions. In the following, first, we will not use any formulation of the Divergence theorem, but we briefly repeat Yavetz's argument \cite{Heav-biblio-Yavez} in order to emphasise how to avoid the use of distributions. Second, we show how Heaviside's specific formula for the cube follows from Gibbs's generalisation of the Divergence theorem, with the help of Heaviside's additional hypothesis, by defining the derivatives in the distribution sense.

In the following, we shall consider scalar functions only. Let $ F\left( \vec{x}\, ,\,\vec{N}(\vec{x})\right)  $ be an arbitrary scalar function, on a closed surface $ \mathcal{S} \in \mathbb{R}^3 $. The closed surface projects into a region $ \mathcal{R} $ in the $ xy-$plane. We assume $ \mathcal{S} $ is vertically simple\footnote{A vertically simple surface $ \mathcal{S} $ is intersected only twice by each vertical line over the interior of $ \mathcal{R} $.}. $ \mathcal{S} $ is then described by two equations, $ z=\alpha (x,y) $ and $ z=\beta(x,y) $, representing the upper and the lower part of the surface respectively. The integral of $ F $ over the surface, namely:
\begin{equation}\label{surf-int}
\int_{\mathcal{S}} F\left( \vec{x}\, ,\,\vec{N}(\vec{x})\right)  dS\; ,
\end{equation}
can be rewritten as a volume integral. Indeed, there will always be some function $ f\left( \vec{x}\, ,\,\vec{N}(\vec{x})\right) $ such that:
\begin{equation}\label{primo}
\int_{\mathcal{S}} F\left( \vec{x}\, ,\,\vec{N}(\vec{x})\right)  dS = \int_{\mathcal{V}} f\left( \vec{x}\, ,\,\vec{N}(\vec{x})\right)  dV\; .
\end{equation} 
Like in the usual proof of the Divergence theorem, we start with the surface integral using the unit versors $ \vec{i} $, $ \vec{j} $ and $ \vec{k} $. The l.h.s. of equation [\ref{primo}] is equivalent to the flux of the vector field $ F\vec{N} $ across the surface, because $\vec{N}\cdot\vec{N} = 1$ by definition on the surface itself, and it can be rewritten as the sum of two integrals, one over the upper surface $ \mathcal{S}_1 $, on which $ z = \alpha(x, y) $ and one over the lower one  $ \mathcal{S}_2 $, on which $ z = \beta(x, y) $, namely:
\begin{eqnarray}
\int_{\mathcal{S}} F\left( \vec{x}\, ,\,\vec{N}(\vec{x})\right)  dS &=& \int_{\mathcal{S}_1} F\left( \vec{x}\, ,\,\vec{N}(\vec{x})\right)  dS + \int_{\mathcal{S}_2} F\left( \vec{x}\, ,\,\vec{N}(\vec{x})\right)  dS \nonumber\\
&=&\int_{\mathcal{S}_1} F\left(\alpha\, ,\,\vec{N}(\alpha)\right)\vec{N}(\alpha)\boldsymbol{\cdot}d\vec{S}+\int_{\mathcal{S}_2} F\left(\beta\, ,\,\vec{N}(\beta)\right)\vec{N}(\beta)\boldsymbol{\cdot}d\vec{S}\nonumber\\
&=&\int_{\mathcal{R}} F\left(\alpha\, ,\,\vec{N}(\alpha)\right)\vec{N}(\alpha)\boldsymbol{\cdot}\vec{n}_\alpha dxdy+\int_{\mathcal{R}} F\left(\beta\, ,\,\vec{N}(\beta)\right)\vec{N}(\beta)\boldsymbol{\cdot}\vec{n}_\beta dxdy\; ,\label{superficie1}
\end{eqnarray}
where we have introduced the following notation $ \vec{N}(\alpha) = \vec{N}(x,y,\alpha (x,y)) $, $ d\vec{S}=\vec{N}dS $, and we have used the outward un-normalised vectors $\displaystyle
\vec{n}_\alpha = \left( \frac{\partial \alpha}{\partial x}\, , \frac{\partial \alpha}{\partial y}\, , 1\right) $ and $\displaystyle
\vec{n}_\beta = \left( -\frac{\partial \beta}{\partial x}\, , -\frac{\partial \beta}{\partial y}\, , -1\right) $ and $ \mathcal{R} $ is the projection of $ \mathcal{S}_1 $ and $ \mathcal{S}_2 $ onto the $ xy $ plane. Let us now suppose that the unit cube is centred in the origin of the coordinates, hence $\displaystyle \alpha(x,y)= z-\frac{1}{2} $ and $ \displaystyle \beta(x,y)= z+\frac{1}{2} $. The normal vector $  \vec{n}_\alpha $ is normalized and then $ \vec{n}_\alpha = \vec{N}_\alpha  $. The integral over the whole closed surface can be obtained by summing the integral over the three coupled opposite squares. The sum along the $ z- $component is equal to eq. [\ref{superficie1}] and by using $ \vec{n}_\alpha = \vec{N}_\alpha  $ the integral reads:
\begin{equation}\label{superficie-cubo-2}
\int_{\mathcal{R}} \left[ F\left(\alpha\, ,\,\vec{k}\right) + F\left(\beta\, ,\,-\vec{k}\right)\right]  dxdy =  \int_{\mathcal{R}} \left[ F\left(\alpha\, ,\,\vec{k}\right) - F\left(\beta\, ,\,\vec{k}\right)\right]  dxdy\; ,
\end{equation}
where we used Heaviside's additional hypothesis. In the last integral, $ \vec{N} $ does not depend on $ \vec{x} $ and therefore can be rewritten in the following equivalent form:
\begin{eqnarray}\label{volume-cubo}
\int_{\mathcal{R}} \left[ F\left(\alpha\, ,\,\vec{k}\right) - F\left(\beta\, ,\,\vec{k}\right)\right]  dxdy &=& \int_{\mathcal{R}}dxdy\int_{\beta(x,y)}^{\alpha(x,y)} \frac{\partial}{\partial z} F\left( \vec{x}\, ,\,\vec{k}\right)  dz\nonumber\\
&=& \int_{\mathcal{V}}  \nabla_3 F\left( \vec{x}\, ,\,\vec{k}\right) dV \; .\nonumber
\end{eqnarray}
Like in Heaviside's original formulation, the derivative has been reconstructed ``by hand'' and no additional hypothesis on the differentiability of the unit normal vector field is needed.
By summing the contribution of all surfaces we get Heaviside's formula:
\begin{equation}
\displaystyle f\left( \vec{x}\, ,\,\vec{N}(\vec{x})\right) = \nabla_1 F\left( \vec{x}\, ,\,\vec{i}\right)  + \nabla_2 F\left( \vec{x}\, ,\,\vec{j}\right)  + \nabla_3 F\left( \vec{x}\, ,\,\vec{k}\right)  . 
\end{equation}

In order to show that Heaviside's formulation is a particular case of the most general version of the Divergence theorem, let us now suppose that $ F\left( \vec{x}\, ,\,\vec{N}(\vec{x})\right)  $ be differentiable in the distribution sense, where $ \vec{N}(\vec{x}) $ is the unit normal vector field of the cube. First, we rewrite the surface-integral [\ref{surf-int}] as follows:
\begin{equation}\label{applico-1}
\int_{\mathcal{S}} F\left( \vec{x}\, ,\,\vec{N}(\vec{x})\right)  dS = \int_{\mathcal{S}} F\left( \vec{x}\, ,\,\vec{N}(\vec{x})\right)\vec{N}(\vec{x})\boldsymbol{\cdot}d\vec{S}\; ,
\end{equation}
because for the cube $ \vec{N}(\vec{x})\cdot \vec{N}(\vec{x}) =1 $ holds again. Second, we apply the generalisation of the Divergence theorem involving dyadics, eq. [\ref{div5}], to eq. [\ref{applico-1}], which reads:
\begin{eqnarray}
\int_{\mathcal{S}} F\left( \vec{x}\, ,\,\vec{N}(\vec{x})\right)\vec{N}(\vec{x})\boldsymbol{\cdot}d\vec{S} &=& 
\int_{\mathcal{V}} \nabla\boldsymbol{\cdot}\left\lbrace F\left( \vec{x}\, ,\,\vec{N}(\vec{x})\right)\vec{N}(\vec{x})\right\rbrace dV\nonumber\\
&=&\int_{[0;1]^3} \left\lbrace \partial_x\left( FN_1\right) +  \partial_y\left( FN_2\right)+ \partial_z\left( FN_3\right)\right\rbrace dxdydz\; ,\label{div-th}
\end{eqnarray}
where we simplified the notation in the last line\footnote{$ \partial_x\left( FN_1\right) $ means $ \nabla_1\left\lbrace  F\left( \vec{x}\, ,\,\vec{N}(\vec{x})\right)N_1(\vec{x})\right\rbrace  $.}. 
In order to show that eq. [\ref{div-th}] is equivalent to Heaviside's Cartesian formula, we focus on the first term and we rewrite it as follows:
\begin{equation}\label{intermedio}
\int_{[0;1]^2}dydz\int_0^1 dx \left\lbrace \partial_x\left( FN_1\right) \right\rbrace = \int_{[0;1]^2}dydz\int_0^1 dx \left\lbrace F(1,y,z,\vec{i}) +  F(0,y,z,-\vec{i}) \right\rbrace  \; ,
\end{equation}
because $ \vec{N}(1,y,z)=\vec{i} $ and  $ \vec{N}(0,y,z)=-\vec{i} $, when $ (y,z)\in [0,1]^2 $. If we use Heaviside's additional hypothesis, i.e. $ F(x,y,z,-\vec{i})=-F(x,y,z,\vec{i}) $, eq. [\ref{intermedio}] reads
\begin{equation}
\int_{[0;1]^2}dydz\int_0^1 dx \left\lbrace F(1,y,z,\vec{i}) -  F(0,y,z,\vec{i}) \right\rbrace = \int_{[0;1]^3} \left\lbrace \partial_x\left( F(x,y,z,\vec{i})\right) \right\rbrace dxdydz\; .
\end{equation} 
By repeating the argument for the second and the third term of eq. [\ref{div-th}], we get Heaviside's Cartesian formula.

\section{Back cover transcription}\label{app1}
\begin{figure}[H]
	\centering
	\includegraphics[width=110mm]{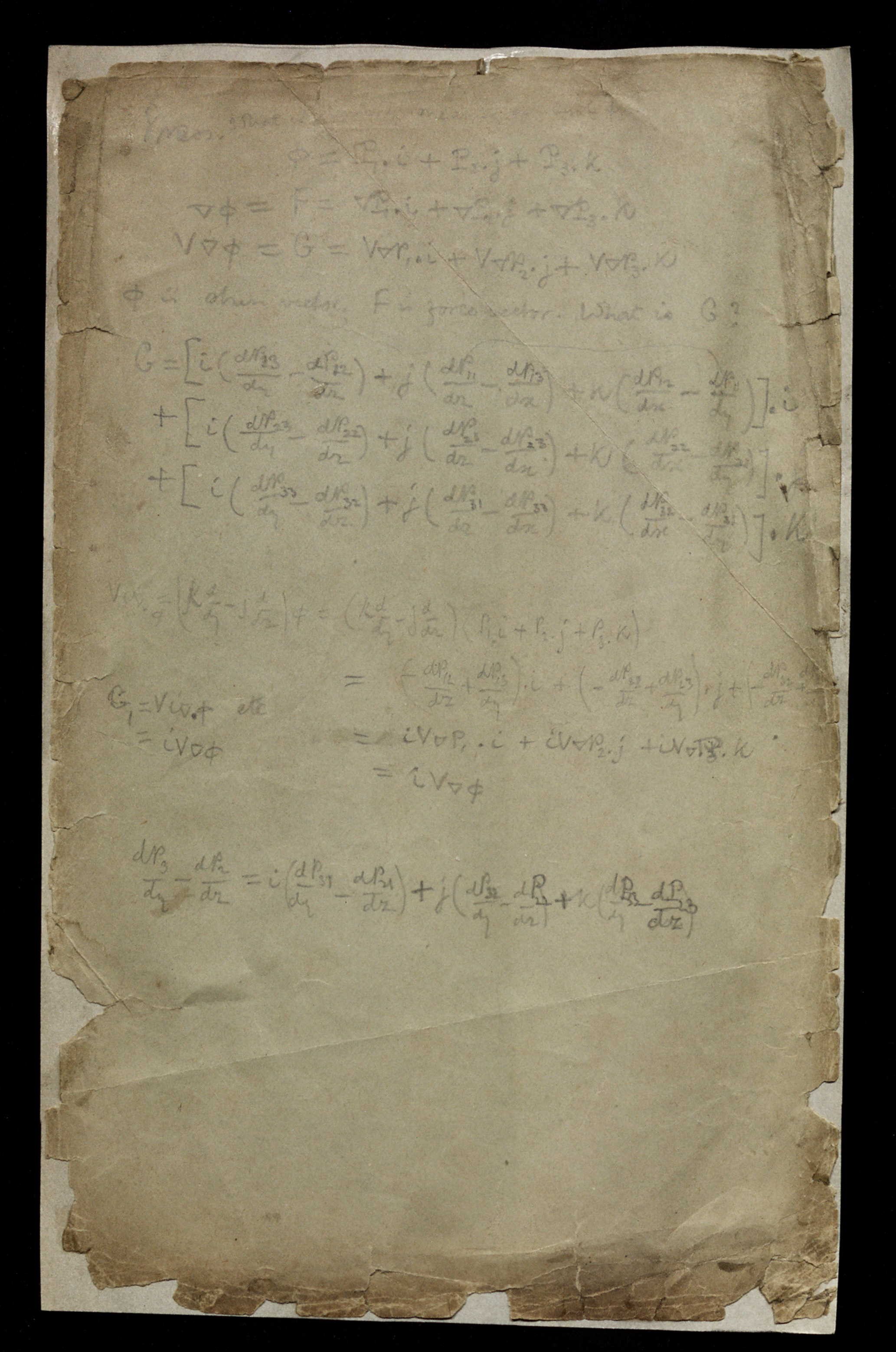}
	\caption{Heaviside's annotation on the back cover of the booklet}
	\label{curl-figure-2bis}
\end{figure} 
Heaviside used $ \phi $ instead of Gibbs's $ \Phi $. In the following, the word in the square bracket cannot be read clearly, because the writing has faded.
\begin{eqnarray}
\nonumber\\
\text{Stress.} &\text{What}& \text{is [the physical] meaning of curl}\; \phi\, ?\nonumber\\
\nonumber\\
&\phi & = \vec{P}_1\, .\,\vec{i} + \vec{P}_2\, .\,\vec{j} + \vec{P}_3\, .\,\vec{k} \nonumber \\
&\nabla\phi & = F = \nabla \vec{P}_1\, .\,\vec{i} +\nabla \vec{P}_2\, .\,\vec{j} +\nabla  \vec{P}_3\, .\,\vec{k} \nonumber\\
& V\nabla\phi & = G =  V\nabla \vec{P}_1\, .\,\vec{i} +V\nabla \vec{P}_2\, .\,\vec{j} +V\nabla  \vec{P}_3\, .\,\vec{k}\nonumber\\
\nonumber\\
\phi\,&\text{is stress} &\text{tensor;}\; F\,\text{is force vector. What is}\; G\, ?\nonumber\\
\nonumber\\
\nonumber\\
G & = & \left[ \vec{i}\left( \frac{dP_{13}}{dy}-\frac{dP_{12}}{dz}\right) +\vec{j}\left( \frac{dP_{11}}{dz}-\frac{dP_{13}}{dx}\right) + \vec{k}\left( \frac{dP_{12}}{dx}-\frac{dP_{11}}{dy}\right) \right] .\; \vec{i} \nonumber\\
& + & \left[ \vec{i}\left( \frac{dP_{23}}{dy}-\frac{dP_{22}}{dz}\right) +\vec{j}\left( \frac{dP_{21}}{dz}-\frac{dP_{23}}{dx}\right) + \vec{k}\left( \frac{dP_{22}}{dx}-\frac{dP_{21}}{dy}\right) \right] .\; \vec{j} \nonumber\\
& + & \left[ \vec{i}\left( \frac{dP_{33}}{dy}-\frac{dP_{32}}{dz}\right) +\vec{j}\left( \frac{dP_{31}}{dz}-\frac{dP_{33}}{dx}\right) + \vec{k}\left( \frac{dP_{32}}{dx}-\frac{dP_{31}}{dy}\right) \right] .\; \vec{k} \nonumber\\
\nonumber\\
\nonumber\\
\nonumber\\
V\vec{i}\nabla\, .\,\phi = \left( \vec{k}\frac{d}{dy}-\vec{j}\frac{d}{dz}\right) \phi &=& \left( \vec{k}\frac{d}{dy}-\vec{j}\frac{d}{dz}\right)\left( \vec{P}_1\, .\,\vec{i} + \vec{P}_2\, .\,\vec{j} + \vec{P}_3\, .\,\vec{k} \right) \nonumber\\
&=& \left( -\frac{dP_{12}}{dz}+\frac{dP_{13}}{dy}\right) .\; \vec{i}+\left( -\frac{dP_{22}}{dz}+\frac{dP_{23}}{dy}\right) .\; \vec{j} + \left( -\frac{dP_{32}}{dz}+\frac{dP_{33}}{dy}\right) .\; \vec{k}\nonumber\\
G =V\vec{i}\nabla .\;\phi\;\;\text{etc}\quad\quad\qquad &=& \vec{i}V\nabla \vec{P}_1 .\;\vec{i} + \vec{i}V\nabla \vec{P}_2 .\;\vec{j} + \vec{i}V\nabla \vec{P}_3 .\;\vec{k}\nonumber\\ 
^{\arrowvert}=\vec{i}V\nabla .\;\phi\;\;\,\,\quad\qquad\qquad &=& \vec{i}V\nabla .\;\phi\nonumber\\
\nonumber\\
\nonumber\\ 
\frac{d\vec{P}_3}{dy} & - & \frac{d\vec{P}_2}{dz}\,\quad =\quad \vec{i}\left( \frac{dP_{13}}{dy}-\frac{dP_{12}}{dz}\right) +\vec{j}\left( \frac{dP_{32}}{dy}-\frac{dP_{22}}{dz}\right) + \vec{k}\left( \frac{dP_{33}}{dy}-\frac{dP_{23}}{dz}\right) \nonumber
\end{eqnarray}

\section{Other annotations by Heaviside}\label{others}
In this appendix, we shall present the annotations we did not discuss in the main text.

\begin{figure}[ht!]
	\centering
	\includegraphics[width=70mm]{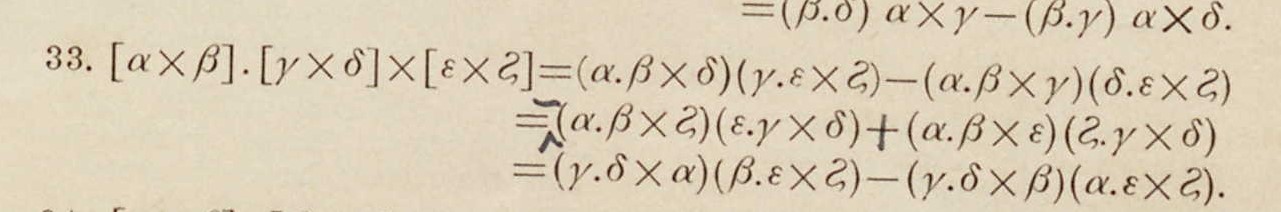}
	\caption{\textbf{Misprint correction.}
		In the third line a minus sign has been added and the minus sign has been changed into a plus sign}
	\label{misprint}
\end{figure}

On pages 6, 8 and 31, three corrected misprints are present. In Fig. \ref{misprint}, we reported one example. 
We suppose that these corrections were made by Gibbs, because they are present also in a different copy of the booklet, conserved by the University of Michigan library. (\cite{Gibbs-pamphlet-2}).

\begin{figure}[ht!]
	\begin{subfigure}{.5\textwidth}
		\centering
		\includegraphics[width=70mm]{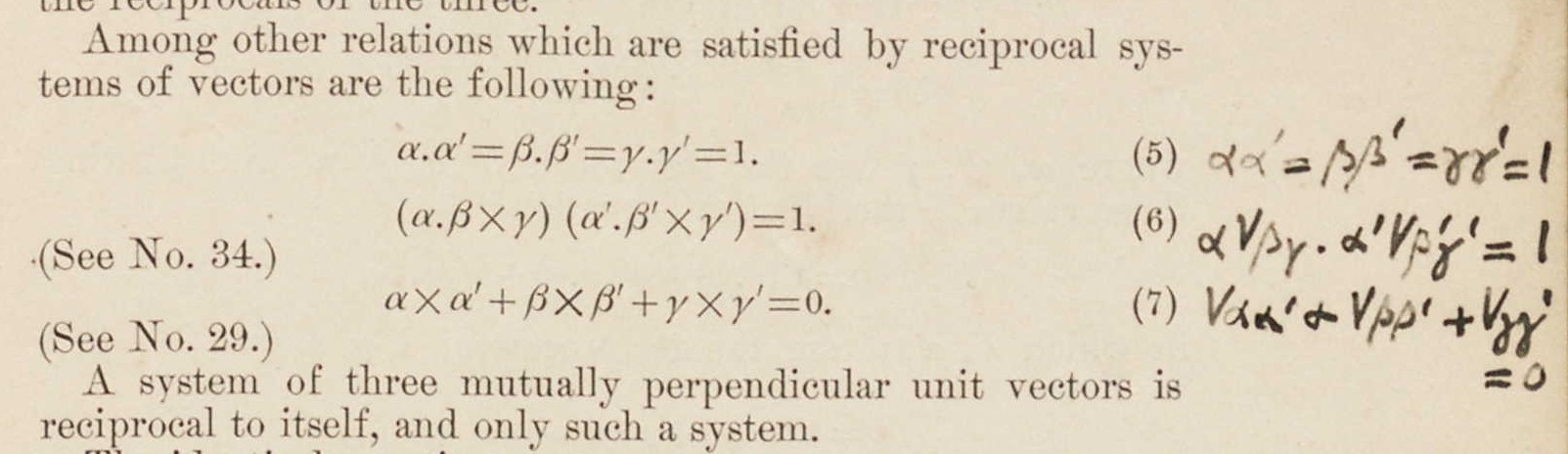}
	\end{subfigure}%
	\begin{subfigure}{.5\textwidth}
		\centering
		\includegraphics[width=65mm]{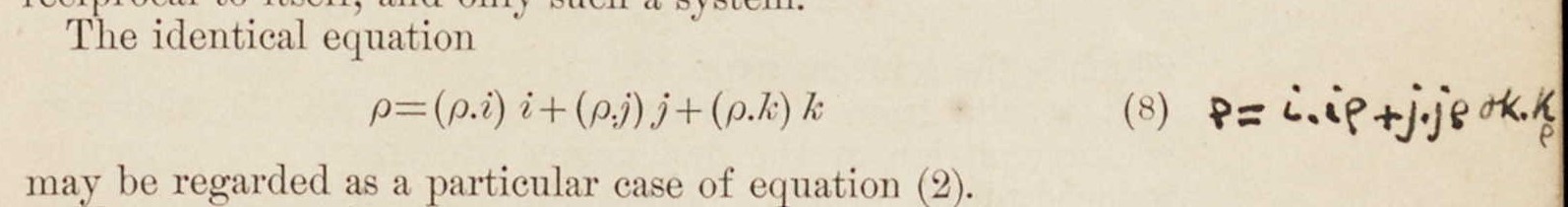}
	\end{subfigure}%
	\caption{Heaviside's translation of Gibbs's equations (5), (6), (7) and (8) on p. 11}
	\label{reciprocal2}
\end{figure}

On the top of page 11, Fig.  \ref{reciprocal2}, Heaviside translated Gibbs's equations (5), (6), (7) and (8).

\begin{figure}[ht!]
	\centering
	\includegraphics[width=90mm]{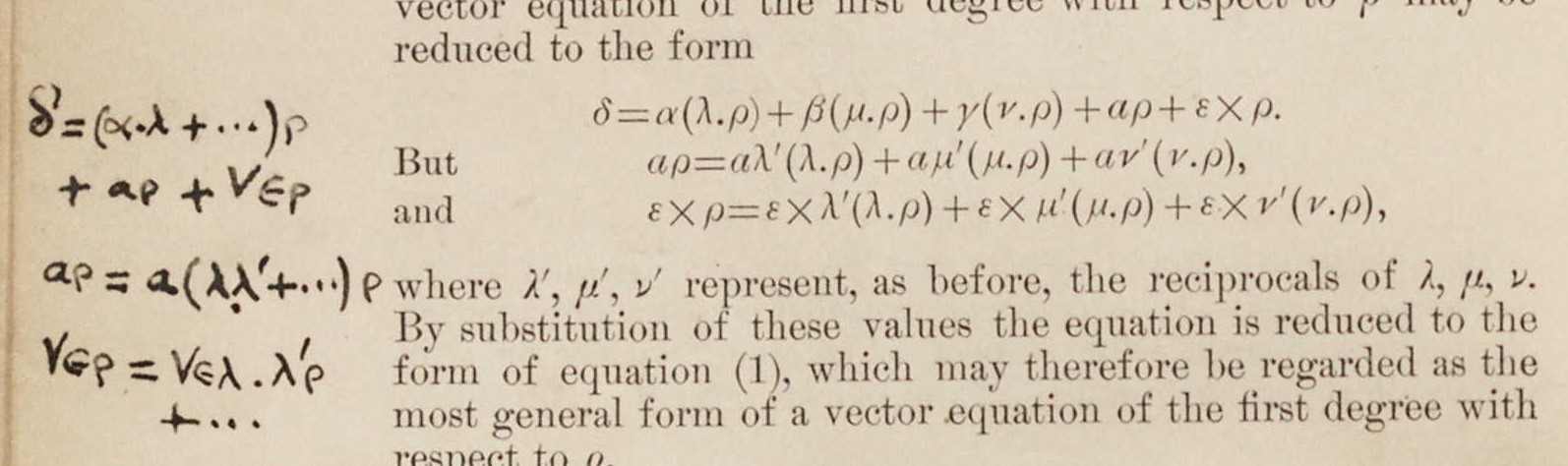}
	\caption{Heaviside's translation on p. 12}
	\label{vector-eq-bis}
\end{figure}

In the left margin of page 12, Heaviside translated an equation regarding vector linear equations, Fig. \ref{vector-eq-bis}.

\begin{figure}[ht!]
	\centering
	\includegraphics[width=90mm]{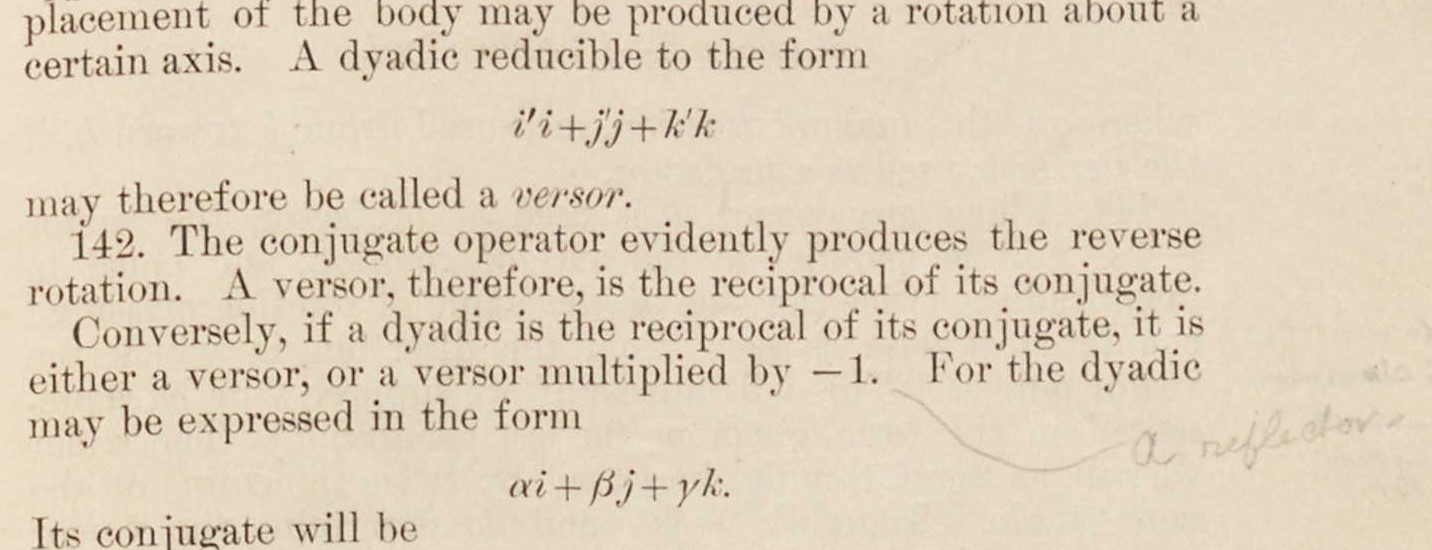}
	\caption{Heaviside's annotation written in pencil: `a reflector' (p. 53)}
	\label{reflector}
\end{figure}

In the last part of chapter 3 in Gibbs's booklet, the author considered rotations and strains as applications of the dyadic's tool. In particular, he remarked how rotations and strain are intimately related to the characteristics of the dyadic itself. In continuum mechanics the linear transformation associated to the dyadic maps the positions of the points of a rigid body subjected to a displacement into a new set of positions, after that stresses or strains have acted on the body itself. If the dyadic is reducible to the form $\mathcal{R} $, see eq. [\ref{rotation-matrix}], the body will suffer no change of form. The converse is also true and Gibbs observed that `the displacement of the body may be produced by a \textit{rotation} about a certain axis.' [emphasis added] (\cite{Gibbs-pamphlet}, 53). This statement clarifies why we chose the letter $ \mathcal{R} $. In his booklet, Gibbs called $ \mathcal{R} $ a \textit{versor}. On the same page, Fig. \ref{reflector}, we found an annotation Heaviside wrote in pencil. 
Gibbs noticed that the conjugate of $ \mathcal{R} $ produces the reverse rotation and from this fact it follows that a versor is the reciprocal of its conjugate. But the converse is not true: if a dyadic is the reciprocal of its conjugate it could correspond either to a versor or to a versor multiplied by $ -1 $. Heaviside's annotation written in pencil shows that he named `reflector' the negative of a versor, see Fig. \ref{reflector}.

\begin{figure}[ht!]
	\centering
	\includegraphics[width=100mm]{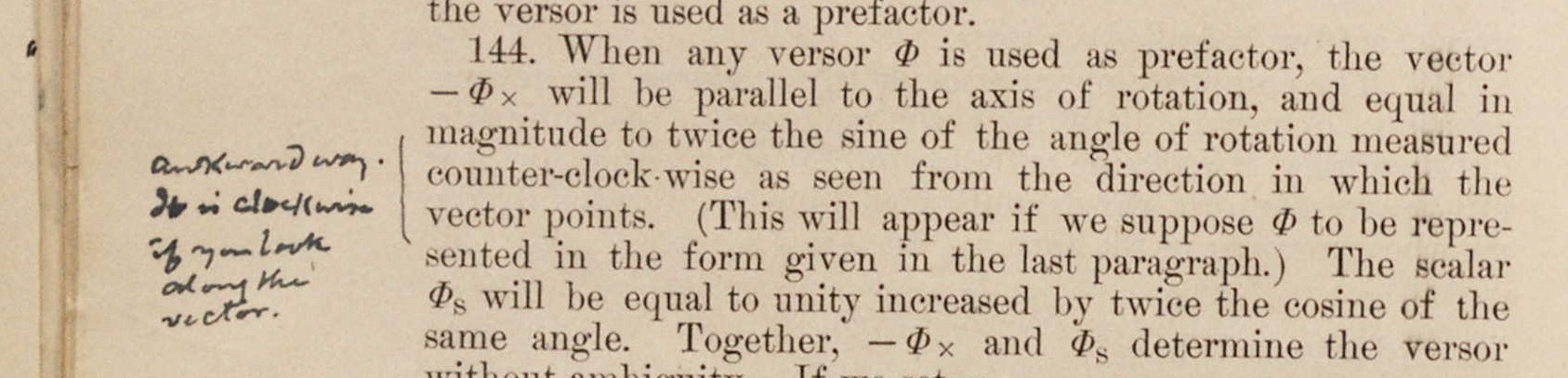}
	\caption{Heaviside's annotation on Gibbs's $ \Phi_{\times} $ (p. 54)}
	\label{phi-cross-comment}
\end{figure}

On page 54, see Fig. \ref{phi-cross-comment}, there is a comment on Gibbs's  $ \Phi_\times  $, i.e. eq. [\ref{phi-cross}]. As already said, Heaviside had introduced it in 1886 by considering displacements of the points of a body due to stresses. He had pointed out that the reversed sign vector, namely $ -\Phi_\times  $, `is required to balance the torque.' (\cite{ELP1}, 544). As we can see in Fig. \ref{phi-cross-comment},
referring to the right-hand reference system of coordinates, Gibbs pointed out that the vector $ -\Phi_{\times} $ `will be parallel to the axis of rotation and equal in magnitude to twice the sine of the angle of rotation \textit{measured counter-clockwise as seen in the direction in which the vector points}.' [emphasis added] (\cite{Gibbs-pamphlet}, 54). Heaviside did not like the emphasised expression and commented: `awkward way it is clockwise if you look along the vector' (\cite{Gibbs-pamphlet} , 54). Of course, there is nothing substantial in Heaviside's annotation, it is just a matter of taste if we choose to look along the vector or from the above.

On pages 56, 57 and 58, there are three generic annotations, two written in pencil and one written in pen, which the author may have written in order to remind himself the topic of a particular formula. 
\begin{figure}[ht!]
	\begin{subfigure}{.33\textwidth}
		\centering
		\includegraphics[width=40mm]{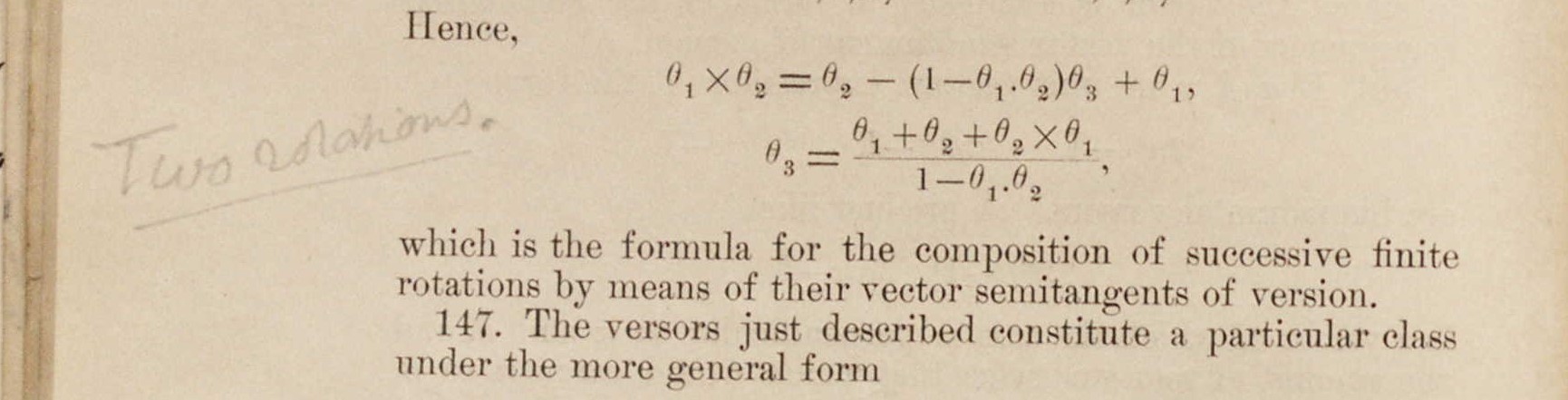}
		\caption{`Two rotations' (p. 56)}
	\end{subfigure}%
	\begin{subfigure}{.33\textwidth}
		\centering
		\includegraphics[width=35mm]{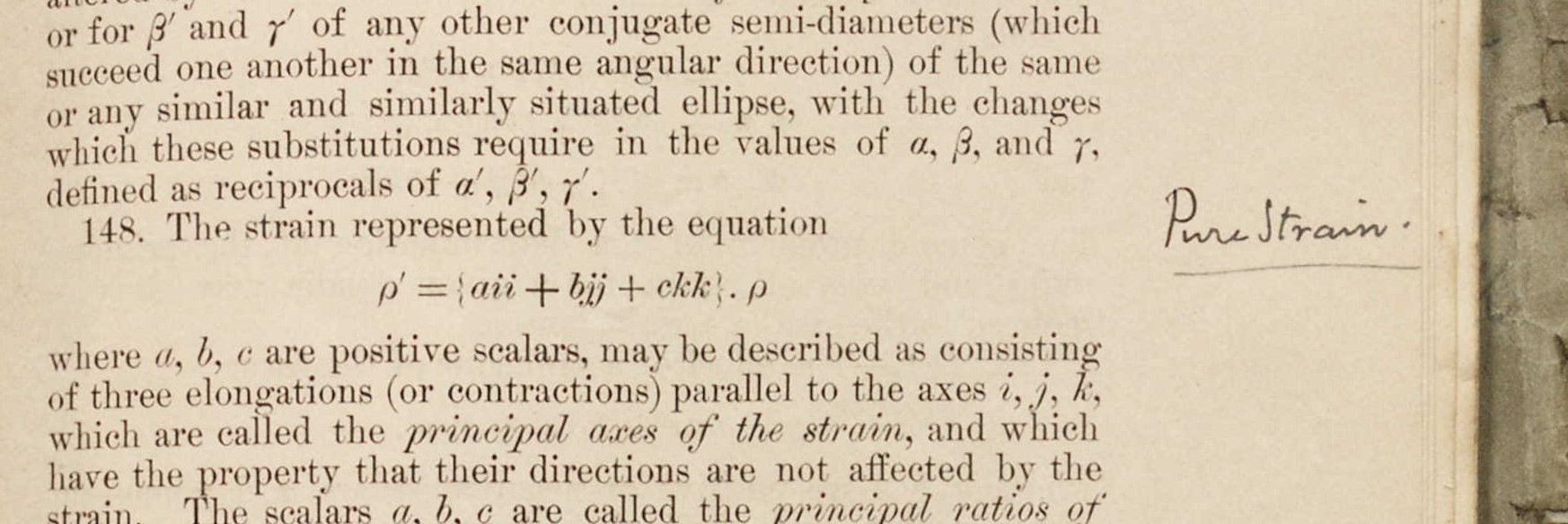}
		\caption{`Pure Strain' (p. 57)}
	\end{subfigure}
	\begin{subfigure}{.33\textwidth}
		\centering
		\includegraphics[width=35mm]{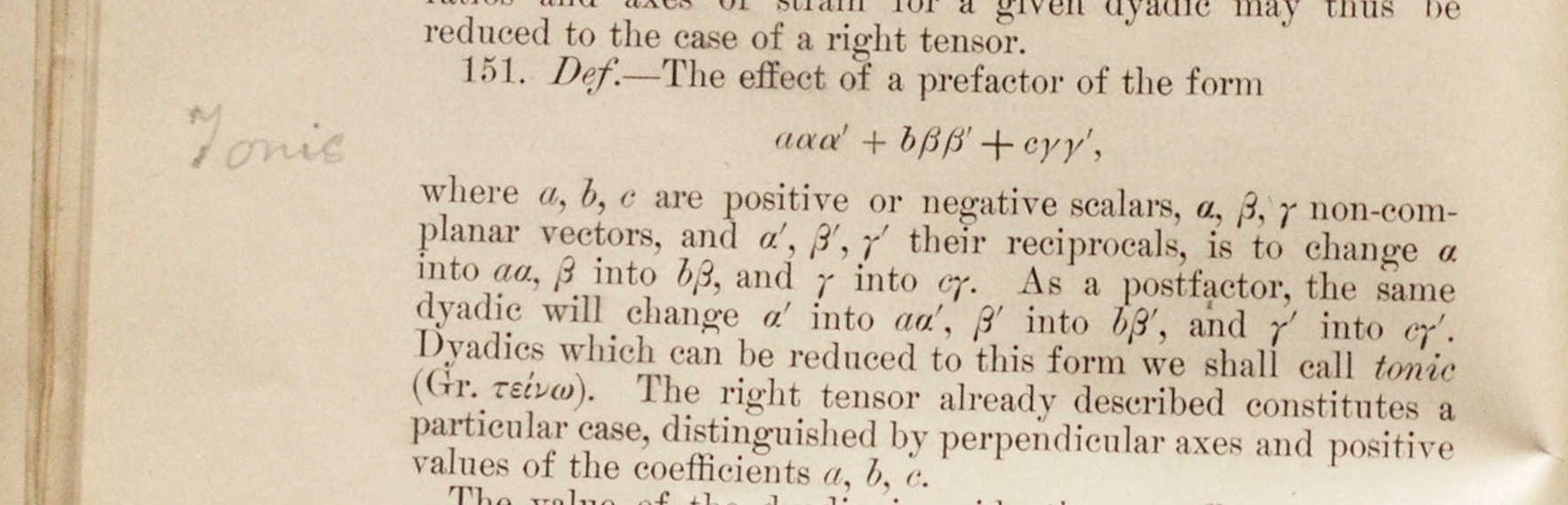}
		\caption{`Tonic' (p.58)}
	\end{subfigure}%
	\caption{Three generic annotations}
	\label{generic}
\end{figure} 
As shown in Fig. \ref{generic}, Heaviside referred to the formula for $\theta_3  $ by writing `Two rotations' (\cite{Gibbs-pamphlet} , 56), while on pages 57 and 58 he simply wrote the terms `Pure Strain' and `Tonic' respectively close to the formula they referred to.

\begin{figure}[ht!]
		\centering
		\includegraphics[width=90mm]{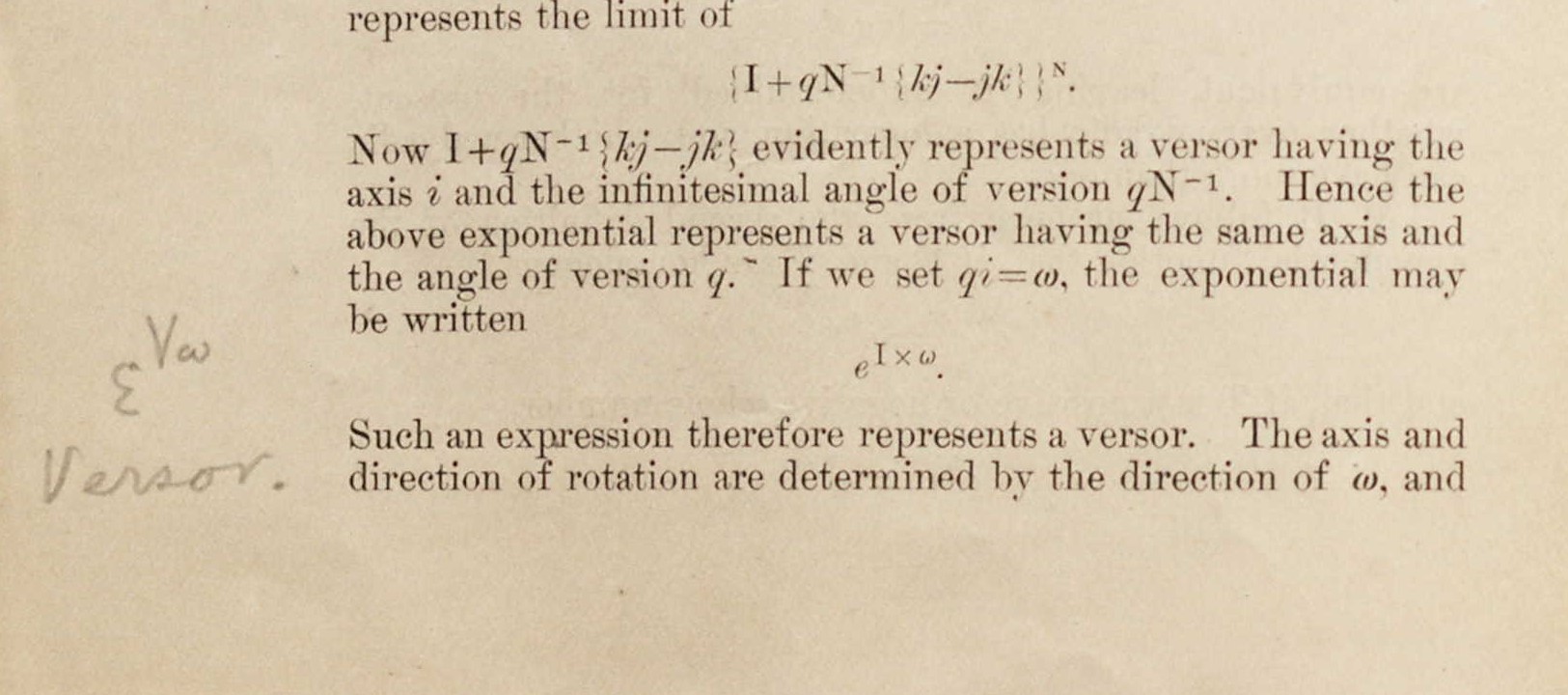}
		\caption{Heaviside's annotation written in pencil: `Versor' (p. 70)}
		\label{versor}
	\end{figure}

On page 70 the versor appears, Fig. \ref{versor}. Here, Gibbs discussed functions of linear operators and Heaviside wrote an annotation in pencil where he reminded himself the meaning of the exponential of the operator $ \mathcal{I}\times\omega $, where $ \mathcal{I} $ has been defined by eq. [\ref{id-dyadic}] and $ \omega $ is a vector. 
\end{appendices}

\end{document}